\documentclass[11pt,a4paper]{article}
\pdfoutput=1
\usepackage{jheppub}
\usepackage[utf8]{inputenc}

\usepackage{color}
\usepackage{amsmath}
\usepackage{mathtools}
\usepackage{esint}
\usepackage{pifont}
\usepackage{bbold}

\usepackage{bbm}
\usepackage{verbatim}   
\usepackage{subfigure}
\usepackage{acronym}

\usepackage{amsfonts}
\usepackage{amssymb}
\usepackage{mathrsfs}
\usepackage{graphicx}
\usepackage{multirow}
\usepackage{slashed}

\usepackage{caption}

\usepackage{soul}

\definecolor{mypurple}{RGB}{164,64,214}

\title{The Weak Scale from Weak Gravity}

\author[a]{Nathaniel Craig,}
\emailAdd{ncraig@physics.ucsb.edu}
\author[b]{Isabel Garcia Garcia,}
\emailAdd{isabel@kitp.ucsb.edu}
\author[a]{and Seth Koren}
\emailAdd{koren@physics.ucsb.edu}

\affiliation[a]{Department of Physics, University of California, Santa Barbara, CA 93106, USA}
\affiliation[b]{Kavli Institute for Theoretical Physics, University of California, Santa Barbara, CA 93106, USA}

\abstract{We explore the prospects for bounding the weak scale using the weak gravity conjecture (WGC), addressing the hierarchy problem by violating the expectations of effective field theory. Building on earlier work by Cheung and Remmen, we construct models in which a super-extremal particle satisfying the electric WGC for a new Abelian gauge group obtains some of its mass from the Higgs, setting an upper bound on the weak scale as other UV-insensitive parameters are held fixed. Avoiding undue sensitivity of the weak scale to the parameters entering the bound implies that the super-extremal particle must lie at or below the weak scale. While the magnetic version of the conjecture implies additional physics entering around the same scale, we demonstrate that this need not correspond to a cutoff for the Higgs potential or otherwise trivialize the bound. We stress that linking the WGC to the weak scale necessarily involves new light particles coupled to the Higgs, implying a variety of experimentally accessible signatures including invisible Higgs decays and radiative corrections in the electroweak sector. These models also give rise to natural dark matter candidates, providing additional paths to discovery. In particular, collective effects in the dark matter plasma may provide a telltale sign of the Abelian gauge group responsible for bounding the weak scale.
}

\begin{document}
	
	\maketitle
	

	\section{Introduction}
	
The problems of the Standard Model (SM) are, for the most part, problems of effective field theory (EFT). As the prospects for resolving these problems in the context of EFT seem increasingly remote, the possibility that their resolution lies in the {\it failure} of EFT becomes ever more compelling. Of course, the robustness of EFT suggests that its failure should only come with good reason, and the challenge remains to identify mechanisms for this to occur without spoiling the remarkable successes of the SM. Ideally, such mechanisms would still leave telltale fingerprints at low energies, making the search for signs of EFT's failure as compelling as the ongoing search for signs of its success.

At the same time, the apparent problems of the SM are tied up in the one force it does not describe: gravity. Given gravity's propensity for connecting the ultraviolet to the infrared, it is natural to look to the weakest force for guidance. Indeed, in the past decade it has become increasingly apparent that a consistent theory of quantum gravity may proscribe the space of viable EFTs. This has been perhaps most clearly articulated by the Swampland program \cite{Vafa:2005ui}, which aims to identify the set of otherwise-consistent effective field theories that cannot be completed into a consistent theory of quantum gravity (see \cite{Brennan:2017rbf, Palti:2019pca} for recent and comprehensive reviews). The potential for the Swampland program, and adjacent developments, to explain some of the more puzzling features of the SM has been explored in e.g.~\cite{ArkaniHamed:2007gg, Dvali:2007hz, Dvali:2007wp, Dvali:2008tq, Cheung:2014vva, Ooguri:2016pdq, Ibanez:2017kvh, Ibanez:2017oqr, Hamada:2017yji, Lust:2017wrl, Gonzalo:2018tpb, Gonzalo:2018dxi, Craig:2018yvw, Kaloper:2019lpl}. 

The Swampland program includes a range of conjectures of varying concreteness, drawing inspiration from both broad infrared considerations and specific properties of ultraviolet completions. In what follows, we will restrict our attention to various forms of the Weak Gravity Conjecture (WGC) \cite{ArkaniHamed:2006dz}, colloquially ``gravity is the weakest force.'' Compared to other conjectures within the Swampland program, the WGC has the advantage of being relatively concrete and motivated by a variety of infrared considerations. As a bound on mass scales, it also provides a suggestive setting for addressing the problems of the SM, most notably the electroweak hierarchy problem.

Although there are many variants of the WGC, we will be primarily interested in its two essential forms. Namely, a $U(1)$ gauge theory with gauge coupling $g$ coupled to gravity must satisfy the following criteria:
\begin{itemize}
\item {\it Electric Weak Gravity Conjecture}: There exists a particle in the theory with mass $m$ and charge $q$ under the $U(1)$ satisfying
\footnote{ Strictly speaking: $m \leq \sqrt{2} g q M_{\rm Pl}$, with $M_{\rm Pl} \equiv 1 / \sqrt{8 \pi G_N} \approx 2 \cdot 10^{18} \ {\rm GeV}$. In the following, we will drop the factor of $\sqrt{2}$ appearing in various forms of the conjecture, as $\mathcal{O}(1)$ corrections to Eq.(\ref{eq:ewgc}) will have no bearing in the subsequent discussion. }	
\begin{equation} \label{eq:ewgc}
	m \lesssim g q M_{\rm Pl} \ ,
\end{equation}
where $g$ is the gauge coupling (i.e.~the electric charge quantum), and $q \in \mathbb{Z}$.
\item {\it Magnetic Weak Gravity Conjecture}: The purely local, electric description of the $U(1)$ breaks down at a scale $\Lambda$ satisfying
\begin{equation} \label{eq:mwgc}
\Lambda \lesssim g M_{\rm Pl} \ .
\end{equation}
\end{itemize}

There are several points worth emphasizing about each of the above conjectures. The first is that we have framed the electric WGC in terms of a particle, making apparent its role in constraining the parameters of EFTs. Some recent attempts to prove the weak gravity conjecture \cite{Cheung:2018cwt, Hamada:2018dde, Bellazzini:2019xts} have largely focused on demonstrating that the bound can be satisfied by states heavier than the Planck scale, such as black holes. While these arguments address some of the rationale for the electric WGC coming from the (in)stability of extremal black holes in non-supersymmetric theories, they are not valid in all regimes and do not broadly accommodate all of the evidence for the WGC. If this were all that there was to the conjecture, then it would be largely trivial, as a weaker version satisfied by trans-Planckian states does not necessarily circumscribe the properties of EFTs embedded in a theory of quantum gravity. In what follows, we will presume (not without good reason) the stronger form of the conjecture involving degrees of freedom in the infrared.

The second point worth emphasizing is that our framing of the electric WGC is somewhat vague about the detailed properties of the particle(s) satisfying the bound -- whether there is more than one, whether they are stable, whether they are the lightest charged particles, whether they are the particles of smallest charge, etc. Various refinements of the electric WGC have been proposed (e.g.~\cite{Heidenreich:2015nta, Heidenreich:2016aqi, Montero:2016tif, Andriolo:2018lvp}), and may give rise to diverse avenues for addressing problems of the SM. For the most part Eq.(\ref{eq:ewgc}) will be sufficient for our purposes, though possible refinements will become relevant when investigating the interplay between the electric and magnetic versions. The electric WGC is also modified in the presence of light (ideally massless) scalar fields \cite{Palti:2017elp}, which potentially allows for separation of the electric and magnetic scales.

The third point worth emphasizing is our identification of the scale $\Lambda$ appearing in the magnetic WGC with the scale at which the purely local, electric description of the $U(1)$ breaks down, rather than a more sweeping cutoff. Although $\Lambda$ is often regarded as the scale at which quantum gravity enters, or four-dimensional QFT breaks down entirely, this need not be the case.
In arguments leading to the magnetic WGC, $\Lambda$ is typically identified with the scale of degrees of freedom linked to the existence of magnetic monopoles.
This could readily be satisfied, for example, if the $U(1)$ in question is the IR remnant of an $SU(2)$ gauge group higgsed by the vacuum expectation value (vev) of a complex scalar transforming in the adjoint representation \cite{tHooft:1974kcl,Polyakov:1974ek}. Provided the vev is at or below the Planck scale, then the scale appearing in Eq.(\ref{eq:mwgc}) can be identified with the appearance of the massive $W$ bosons. Of course, there are more radical possibilities for physics at the scale $\Lambda$ -- see e.g.~\cite{Heidenreich:2017sim} for an extended discussion -- but it is important to bear in mind that arguments leading to the magnetic conjecture may be satisfied without leading to the end of four-dimensional QFT.
We emphasize that this is not in contradiction with statements made in the context of stronger versions of the WGC that indicate a cutoff to four-dimensional QFT at scales of order $g^{1/3} M_{\rm Pl}$ \cite{Heidenreich:2016aqi,Heidenreich:2017sim} associated to the appearance of an infinite tower of particles: In the regime of weak coupling $g \ll 1$, and therefore the cutoff $g^{1/3} M_{\rm Pl}$ lies parametrically above the scale $g M_{Pl}$.

In this paper, we will explore the prospects for bounding the value of the weak scale using the electric WGC, thereby addressing the electroweak hierarchy problem. Such efforts must necessarily also contend with the magnetic WGC, which implies physics appearing around the same scale, and that may trivialize the purported bounds. In doing so, our work builds on \cite{Cheung:2014vva}, where the first such attempts were made, and \cite{Lust:2017wrl}, which attempted to separate the scales implied by electric and magnetic forms of the conjecture. This approach is broadly inspired by earlier work involving discrete symmetries \cite{Dvali:2007hz, Dvali:2007wp, Dvali:2008tq, Craig:2018yvw}, and discrete versions of both the model in \cite{Cheung:2014vva} and the models presented here were first sketched in \cite{Dvali:2008tq}, albeit with significantly different implications. Needless to say, our attempts to bound the weak scale with the weak gravity conjecture are also inspired by the application of other Swampland conjectures to the same problem \cite{Ooguri:2016pdq, Ibanez:2017kvh, Ibanez:2017oqr, Hamada:2017yji, Gonzalo:2018tpb, Gonzalo:2018dxi}.

To see how the WGC might be linked to the hierarchy problem, it is worth briefly reviewing the logic behind the problem itself. In an EFT with UV-sensitive parameters (such as the Higgs mass in the SM), the natural expectation is that these will accumulate contributions from physics at and above the cutoff $\Lambda_{\rm EFT}$ of the EFT. Absent substantial cancellations between these contributions (which must be considered fine-tunings in the absence of special symmetries or dynamics, since decoupling implies that physics at $\Lambda_{\rm EFT}$ is insensitive to the far infrared), the natural size of a UV-sensitive parameter is $\mathcal{O}(1)$ in units of $\Lambda_{\rm EFT}$. Viewed as a statement about probability distributions in the space of EFTs, this amounts to saying that the most likely EFTs are the ones whose UV-sensitive parameters are on the order of (appropriate powers of) the cutoff. Smaller values of such parameters can be explained by reference to symmetries, dynamics, or anthropics, which all have the effect of shaping the probability distribution on the space of theories so that it peaks at smaller values of the parameter of interest -- in some cases by lowering the cutoff itself, in others by enforcing seemingly unlikely cancellations. The WGC can operate in the same way: given a consistent theory of quantum gravity in the UV, WGC bounds proscribe the space of EFTs by bounding the masses of certain charged particles. If the mass of a particle subject to a WGC bound can be made to depend on the value of the weak scale, then the WGC will shape the probability distribution for the weak scale. The resulting probability distribution may peak at values much smaller than naive EFT reasoning would suggest, thereby providing a rationale for why $v \lll \Lambda_{\rm EFT}$.
The observed value of the weak scale in this case might result from seemingly-arbitrary cancellations between UV contributions and thus appear fine-tuned to an infrared observer, but it would be ``natural'' in the sense that it represents
a likely value in the distribution. Put another way, satisfying the WGC bound would orchestrate correlations among UV parameters that might appear surprising from an EFT perspective.

Of course, the space of EFTs is vast, even if the low-energy field content is held fixed: there are many parameters. In this parameter space, WGC bounds may be satisfied by simultaneously varying the values of both UV-sensitive and -insensitive parameters, creating flat directions in which all values of the former are allowed. To effectively bound the weak scale using the WGC, we must therefore restrict ourselves to slices of the space in which UV-insensitive parameters are held fixed while UV-sensitive parameters are allowed to vary. While this might seem like a radical requirement, it is really not so different from more conventional solutions to the hierarchy problem. In supersymmetric or composite extensions of the SM, the Higgs mass parameter (and hence the weak scale) is fixed by technically natural parameters that may be rendered small by separate mechanisms. Dynamical approaches such as the relaxion \cite{Graham:2015cka} likewise relate dynamically-selected values of the weak scale to fixed values of technically natural parameters. The anthropic bound on the weak scale following from the atomic principle \cite{Agrawal:1997gf} requires holding the dimensionless parameters of the SM fixed across the landscape of vacua. In each case, the technically natural (or otherwise UV-insensitive) parameters can be set by other considerations, and the probability distribution on the space of EFTs restricted to those that are UV-sensitive.

Although the value of the weak scale resulting from a WGC bound may appear fine-tuned from an infrared perspective, there is still a notion of naturalness in the EFT relating the values of technically-natural parameters to the value of the weak scale. If the particle subject to a super-extremality bound only obtains a small fraction of its mass from electroweak symmetry breaking, and a much larger part from technically natural parameters, then the value of the weak scale saturating the WGC bound is inordinately sensitive to the technically natural parameters: small fractional variations in these parameters lead to large fractional variations in the value of the weak scale saturating the WGC bound. If the WGC plays a role in fixing the weak scale, this suggests that the physics implied by the electric and magnetic versions of the conjecture should be near or below the weak scale. In this respect, a WGC bound on the weak scale is not unlike more conventional approaches to the hierarchy problem, although the detailed signatures are radically different. 
 
There are many ways in which this line of reasoning could break down. The scale implied by the magnetic WGC could also be the cutoff scale of the Higgs potential (e.g.~the scale of supersymmetry, or the scale of breakdown of four-dimensional effective field theory), in which case the conventional picture of naturalness is restored without any particular involvement of the conjecture in stabilizing the weak scale. Similarly, the cancellations between large UV contributions required by a WGC bound could be enforced by the appearance of new light particles to do the trick, such as the relaxion. Alternatively, new degrees of freedom appearing at the magnetic scale could themselves satisfy the electric version, relieving the apparent WGC particle of its duties and thereby trivializing a putative bound on the weak scale. More surprisingly, it may be that UV-sensitive parameters actually control the size of apparently UV-insensitive ones, so that it is no longer sensible to use fixed values of the latter to bound the former. Indeed, this is precisely what appears to happen for gauge fields that emerge as a low-energy description of a scalar field theory: the strength of the emergent gauge coupling is controlled by the scalar masses themselves \cite{Harlow:2015lma}. Whether any of these occurs depends on the UV completion, and different UV completions presumably satisfy the various forms of the WGC in different ways. Our goal here is to determine, from the bottom up, whether satisfying the electric WGC might lead to a bound on the weak scale that is not immediately superseded or trivialized by considerations stemming from the magnetic version.

Even if the WGC can be successfully linked to the weak scale, it bears remembering that the most satisfying solutions to the hierarchy problem do not merely bound the value of the weak scale, but also explain why electroweak symmetry is broken in the first place. Although the models presented here may explain why $v$ is not large, they do not address why it is nonzero. Of course, this is not a fatal shortcoming; large positive values of the Higgs mass-squared could be disfavored by additional considerations including properties of the UV completion (e.g. frameworks that favor electroweak symmetry breaking, such as supersymmetry, or details of the physics at $\Lambda$), additional WGC bounds coming from multiple $U(1)$s, scanning or other cosmological evolution that can induce a bias, anthropics, and so forth.

Given these caveats, a WGC bound on the weak scale without testable predictions would be little more than a way to help theorists sleep at night, and theorists are not entitled to much sleep in the current era. The value of the models presented here is that they make relatively sharp predictions for the mass scale and couplings of the new particles that connect the WGC to the weak scale, providing a motivated target for experimental searches. Falsifying these models would not rule out the WGC or its potential role in stabilizing the weak scale, but would certainly narrow the path for it to do so.

This paper is organized as follows: In section \ref{sec:weakscale} we present a series of models that connect the WGC bound to the weak scale.
We begin by reviewing the gauged $U(1)_{B-L}$ model proposed in \cite{Cheung:2014vva}.  While perhaps the simplest means to our end, this model requires the magnetic WGC to be addressed well below the weak scale, in conflict with experimental bounds. This leads us to consider models with a new $U(1)$ gauge group and a WGC particle that acquires at least some of its mass from electroweak symmetry breaking. We consider models in which the WGC particle is alternately a scalar or a fermion. In both cases, avoiding additional fine-tuning implies that these new particles must couple to the Higgs and lie at or below the weak scale. We further discuss how the magnetic version of the conjecture may be accommodated in these models.
In section \ref{sec:scalarWGC} we explore the possibility of parametrically separating the mass of the WGC particle from the magnetic scale due to the effects of Yukawa forces. We review the examples of scale separation presented in \cite{Lust:2017wrl} and adapt them to our models, finding a degree of tension between scale separation and additional fine-tuning. Successfully linking the WGC to the weak scale necessarily involves new particles coupled to the Higgs, giving rise to a variety of signatures that we explore in section \ref{sec:pheno}. Our models give rise to natural dark matter (DM) candidates that further shape the motivated parameter space, and provide additional paths to discovery. In particular, collective effects in the DM plasma may provide a probe of exceedingly small values of the gauge coupling in the context of galaxy cluster collisions, regardless of its relation to the hierarchy problem.

\section{The weak scale from weak gravity}
\label{sec:weakscale}

The basic observation that the upper bound set by the WGC on the mass of a state charged under an unbroken $U(1)$ could, indirectly, translate into a bound on the vev of a scalar field is not new \cite{Cheung:2014vva}.
In its simplest form, a Dirac fermion with mass $m$ and Abelian charge $g$ satisfying the super-extremality bound must have $m \lesssim g M_{\rm Pl}$. If the fermion mass arises from a Yukawa coupling to a scalar field that gets a non-zero vev $f$, of the form $m \sim y f$, then the WGC bound translates into a bound on $f$, of the form
\begin{equation}
	\frac{f}{M_{\rm Pl}} \lesssim \frac{g}{y} \ .
\label{eq:WGCtoy}
\end{equation}
The size of a UV-sensitive parameter is therefore bounded by two technically-natural couplings.
Of course, an upper bound $f / M_{\rm Pl} \ll 1$ would only follow from Eq.(\ref{eq:WGCtoy}) for equally small values of $g / y$. To retain perturbativity we must have $g ,y \lesssim 1$, and thus such a small ratio would require exceedingly small values of the gauge coupling.
In this context, the question of why $f / M_{\rm Pl}$ is small therefore translates into why $g$ itself is tiny.
The crucial observation being, of course, that $g$ and $y$ are UV-insensitive parameters that depend at most logarithmically on the cutoff, and so the qualitative nature of the problem is radically altered.

In particular, the upper bound on $f$ given by Eq.(\ref{eq:WGCtoy}) may be parametrically below the EFT cutoff, which in principle could be as high as $M_{\rm Pl}$.
As discussed in the Introduction, this result would seem to contradict EFT logic, but can nevertheless be reconciled with the notion of naturalness when extra consistency conditions believed to be necessary for UV completion are taken into account -- namely, that the WGC must hold. In this sense, the bound on $f$ set by the WGC would `explain away' what from the EFT perspective looks like a hideous tuning, making it in fact natural for the ratio $f/M_{\rm Pl}$ to be as small as $\sim g / y$.
Of course, values of the vev such that $f / M_{\rm Pl} \ll g / y$ would not be justified by Eq.(\ref{eq:WGCtoy}), and, in the absence of any explanation or dynamical mechanism, an additional hierarchy would demand fine-tuning in the traditional sense.

The difficulty lies in adapting the observation sketched above to obtain an upper bound on the weak scale, and requires introducing an Abelian gauge group with coupling $g$ not bigger than $v / M_{\rm Pl} \sim 10^{-16}$.
We discuss previous attempts at implementing the above discussion by means of a gauged $U(1)_{B-L}$ symmetry in section \ref{sec:noBminusL}, and summarise the roadblocks this suggestion encounters.
Motivated by such difficulties, we present two alternative models in sections \ref{sec:scalar} and \ref{sec:fermion}, where the super-extremal states charged under a new Abelian gauge group (different from $U(1)_{B-L}$) are respectively a scalar and a fermion. Although less minimal in their field content than the suggestion summarised in \ref{sec:noBminusL}, they are nevertheless successful in explaining the smallness of the weak scale.
Finally, section \ref{sec:magneticWGC} discusses how the magnetic version of the conjecture might be addressed in the context of the models introduced in \ref{sec:scalar} and \ref{sec:fermion} without altering the bound on the weak scale.

\subsection{The trouble with $B-L$}
\label{sec:noBminusL}
	
A priori, the most economical approach to try and relate the WGC bound on the mass of a state charged under an Abelian gauge symmetry to the size of the weak scale would be through a gauged, unbroken $U(1)_{B-L}$, as first suggested in \cite{Cheung:2014vva}.
Current experimental constraints from fifth force searches place stringent upper bounds on the corresponding gauge coupling $g \lesssim 10^{-24}$ \cite{Adelberger:2009zz,Wagner:2012ui}. As a result $g M_{\rm Pl} \lesssim 1 \ {\rm keV}$, leaving neutrinos as the only SM particles eligible to satisfy the super-extremality condition.
In fact, that neutrinos saturate the WGC would require $g \gtrsim m_\nu / M_{\rm Pl} \sim 10^{-28}$, a lower bound on the corresponding gauge coupling only four orders of magnitude below current constraints.
	
An unbroken $U(1)_{B-L}$ can be reconciled with the need for neutrino masses so long as they are Dirac particles. Extending the SM field content to include right-handed neutrinos, $\nu^c$, a non-zero mass can be written through a standard Yukawa coupling to the Higgs of the form $\mathcal{L} \sim y_\nu H L \nu^c$, leading to $m_\nu \sim y_\nu v$. The super-extremality constraint mandated by the WGC then translates into an upper bound on the Higgs vev, of the form \cite{Cheung:2014vva}
\begin{equation}
	\frac{v}{M_{\rm Pl}} \lesssim \frac{g}{y_\nu} \ .
\label{eq:neutrinoWGC}
\end{equation}
A Yukawa coupling $y_\nu \sim 10^{-12}$ would then imply an upper bound on the Higgs vev of $\sim 1 \ \rm{TeV}$, a priori providing an explanation for the apparent fine-tuning of the weak scale.

A full resolution would further require addressing the implications of the magnetic form of the conjecture, which, as reviewed in the Introduction, imply some form of cut-off at the scale $\Lambda \lesssim g M_{\rm Pl}$ related to the structure of magnetic monopoles. One possibility would be to embed the $U(1)_{B-L}$ gauge group into an $SU(2)_{B-L}$, broken down to its Abelian subgroup by the non-zero vev of a Higgs field transforming under the adjoint representation. New degrees of freedom would indeed appear below the scale $g M_{\rm Pl}$, these being the $W$ gauge bosons corresponding to the broken directions, and carrying 2 units of $U(1)_{B-L}$ charge. For a natural value of the adjoint vev close to the Planck scale, these new states would have masses of order $m_W \sim 0.1 \ {\rm eV}$, a possibility that is consistent with current experimental constraints given their exceedingly small coupling to SM degrees of freedom.
	
However, difficulty arises when we try to embed the SM fermion content into appropriate $SU(2)_{B-L}$ multiplets. The fact that the ratio of $B-L$ charges between leptons and quarks is 3, together with the requirement of maintaining gauge anomaly cancellation, implies that each SM fermion field must transform \emph{at least} under a $\bf 4$ of $SU(2)$.
Under the unbroken $U(1)_{B-L}$ subgroup, the different components of a $\bf 4$-multiplet carry charges $-3/2,-1/2, +1/2$, and $+3/2$ (in units of the non-Abelian gauge coupling), one of which can always be identified with the corresponding SM quark or lepton. The remaining 3, however, would correspond to extra degrees of freedom carrying the same SM quantum numbers as the corresponding fermion. Effectively, this would imply the existence of no less than 9 extra fermion families, a possibility which is obviously ruled out.

One could try to embed both the $U(1)_{B-L}$ and the SM gauge sector into the same non-Abelian gauge group, as it is common in the context of GUTs \cite{Slansky:1981yr}. Such an arrangement would avoid the proliferation of copies of the SM fermion families, but it would lead to a $B-L$ gauge coupling $g_{B-L} = \mathcal{O}(1)$, in flagrant violation of experimental constraints. In this context, the $U(1)_{B-L}$ would need to be broken at a high scale. Of course, completion into a non-Abelian group is just one possibility for the physics at $\Lambda$, but it seems unlikely that other options (the appearance of extra dimensions, breakdown of local field theory, etc.)~would be in better agreement with data, given that the SM is well understood at energies far above an eV.
	
The above observations raise the question of whether the presence of an unbroken $U(1)_{B-L}$ gauge symmetry, although possible \emph{in principle}, may not be a viable option \emph{in practice} -- that is when trying to implement a model that addresses the structure of magnetic monopoles while remaining compatible with all experimental constraints.
(We note that trying to play `tricks' such as those in \cite{Saraswat:2016eaz} won't help: whereas a $U(1)_{B-L}$ `clockwork' construction would provide a dynamical explanation of the smallness of $g$ while exponentially deferring the WGC cut-off, the same issues as outlined above (as well as others) will have to be faced when trying to implement a model that addresses the structure of the magnetic monopoles charged under the multiple $U(1)$ factors that comprise the clockwork construction.) From the model-building perspective this provides an interesting direction for further investigation, whereas on the experimental front it reinforces the important role of fifth force searches.

Relating the WGC bound on a massive particle to the electroweak scale thus seems to require the introduction of a new Abelian gauge group with charged matter that acquires at least some of its mass from the Higgs.
In the remainder of this section we present two concrete suggestions that successfully realise this idea.


\subsection{A scalar model}
\label{sec:scalar}

The simplest model that successfully relates the WGC bound on a particle's mass to the size of the electroweak scale only requires the addition of a complex scalar field $\Phi$ charged under a new Abelian gauge group $U(1)_X$, but otherwise being a SM-singlet. Using a WGC bound on the mass of a scalar to constrain the weak scale is a bit subtle, as its mass-squared parameter is not itself a technically natural quantity, but it provides the clearest illustration of the phenomenon, subject to some caveats.

For simplicity, we take $\Phi$ to carry the unit of $U(1)_X$ charge $g$. At the renormalizable level, the scalar potential for $\Phi$ admits the following marginal and relevant interactions:
\begin{equation}
	- \mathcal{L} \supset m^2 |\Phi|^2 + \lambda |\Phi|^4 + \kappa |\Phi|^2 |H|^2 \ ,
\label{eq:scalar}
\end{equation}
and we will focus on the case $\kappa > 0$.
After electroweak symmetry breaking, the physical mass-squared of the complex scalar is given by
	\begin{equation}
	m_\Phi^2 = m^2 + \frac{\kappa}{2} v^2 \ .
	\label{eq:mPhi2}
	\end{equation}
If $m_\Phi^2 < 0$, $\Phi$ acquires a vacuum expectation value and Higgses $U(1)_X$, in which case there is no strong bound from the electric WGC (for possible exceptions, see \cite{Craig:2018yvw}).
However, if $m_\Phi^2 > 0$, requiring that $\Phi$ satisfies the electric form of the conjecture implies
	\begin{equation}
	m_\Phi \lesssim g M_{\rm Pl} \ ,
	\label{eq:scalarWGC}
	\end{equation}
and, in turn, an upper bound on the weak scale as given by
	\begin{equation} \label{eq:scalarbound}
	v^2 = \frac{2}{\kappa} (m_\Phi^2 - m^2) \lesssim \frac{2}{\kappa} \left( (g M_{\rm Pl})^2 - m^2 \right) \, .
	\end{equation}

Turning Eq.(\ref{eq:scalarbound}) into a meaningful upper bound on the size of the weak scale is slightly more subtle than in the toy example discussed at the beginning of section \ref{sec:weakscale}.
Now, $m_\Phi^2$ receives an extra contribution from the mass-squared of $\Phi$, which is itself a UV-sensitive parameter. From an EFT perspective, the natural size for the mass-squared and quartic coupling of $\Phi$ is of the form (up to logarithms) $m^2 \sim \xi \Lambda_{\rm EFT}^2 / 16 \pi^2$, and $\lambda \sim \xi^2 / 16 \pi^2$, with $\xi = \kappa, g^2$, whereas $m_H^2 \sim y_t^2 \Lambda_{\rm EFT}^2 /16\pi^2$ for the SM Higgs. \footnote{These expectations all assume the leading possibility $m^2 \sim \Lambda_{\rm EFT}^2$ is forbidden by a symmetry at $\Lambda_{\rm EFT}$.}
Eq.(\ref{eq:scalarWGC}) could then be satisfied even for $\kappa v^2, |m^2| \gg (g M_{\rm Pl})^2$, provided an accurate cancellation takes place between the two terms in Eq.(\ref{eq:mPhi2}).
The WGC would then explain the apparent fine-tuning in the physical mass of $\Phi$, but the corresponding upper bound would nevertheless be consistent with an electroweak symmetry breaking scale close to the cutoff.

Some further extra assumption needs to be made in order to obtain a meaningful bound on $v$. Ideally, $m^2$ would be itself a technically natural parameter, as would happen if, for instance, $\Phi$ were part of an (approximate) supersymmetric sector, or if $\kappa, g \lll 1$.
In such case, $m^2$ could be consistently argued to be small, and Eq.(\ref{eq:scalarbound}) would imply, parametrically
\begin{equation}
	\frac{v}{M_{\rm Pl}} \lesssim \frac{g}{\sqrt{\kappa}} \ .
\end{equation}
A specific model with $g / \sqrt{\kappa} \sim 10^{-16}$ would then suggest that $v / M_{\rm Pl} \lesssim 10^{-16}$, providing an explanation for the smallness of this ratio.
A prediction of this scenario would be the presence of a complex scalar field $\Phi$ appearing at the scale $m_\Phi \sim \sqrt{\kappa} v$, which may be at or below the weak scale, depending on the value of $\kappa$, and carrying charge $g \lll 1$ under a new Abelian gauge group. Some potentially interesting phenomenological aspects of such a prediction are explored in section \ref{sec:pheno}.

A further observation is that Eq.(\ref{eq:scalarbound}) would seem to allow us to set a non-trivial bound on $v^2$ while taking both $m_\Phi^2 \sim (g M_{\rm Pl})^2, m^2 \gg v^2$. This would decouple $\Phi$ from the Higgs sector, thereby stabilizing the weak scale without leaving any evidence in the infrared.
That this realization would be fine-tuned, in the traditional sense, is evident from the right-hand-side of Eq.(\ref{eq:scalarbound}), where an accurate cancellation would be required. It can also be more quantitatively expressed in terms of the sensitivity of the weak scale to variations in the underlying parameters, precisely as we do for more conventional approaches to the hierarchy problem.\footnote{The rationale for considering the sensitivity of the weak scale to the values of infrared parameters (but not ultraviolet parameters) bears emphasizing. Satisfying the WGC enforces correlations among UV contributions to the mass of the WGC particle, but we should still ask that the value of the weak scale not change radically under small variations in the WGC bound or the parameters connecting the two.}
In particular, we can compute the sensitivity to dimensionful inputs using the Barbieri-Giudice measure \cite{Barbieri:1987fn}:
	\begin{equation}
	\Delta_x \equiv \left| \frac{\partial \log v^2}{\partial \log x} \right| \ .
	\end{equation}
	Treating $m_\Phi^2$ (equivalently, $g$) and $m^2$ as input parameters, the sensitivity of the weak scale to these inputs is then simply
	\begin{equation}
		\Delta_{m_\Phi^2} = \frac{2 m_{\Phi}^2}{\kappa v^2} \ , \qquad {\rm and} \qquad \Delta_{m^2} = \frac{2 m^2}{\kappa v^2} \ .
	\end{equation}
	Indeed, having $m^2, m_\Phi^2 \gg \kappa v^2$ would thus lead to $\Delta_{m^2}, \Delta_{m_\Phi^2} \gg 1$, signalling the mandatory fine-tuning.
	Using the WGC bound on a scalar particle to explain the smallness of $v$, while at the same time avoiding tuning of the infrared parameters, thus requires $\kappa v^2 \sim m^2 \sim (g M_{\rm Pl})^2$.
	Note that the UV-sensitivity of $m^2$ naively implies $v \sim \Lambda_{\rm EFT}/4 \pi$, which is only smaller than the SM expectation by a factor of $\sim y_t / \sqrt{\lambda_{\rm SM}}$. Amusingly, this is still the same improvement one might expect from far more elaborate models of neutral naturalness \cite{Contino:2017moj}, but in this case it is primarily a casualty of assigning the same cutoff scale to $\Phi$ and $H$. In a supersymmetric UV completion, the one-loop correction to $m^2$ induced by $\kappa$ would be cut off by the Higgsino mass, leading to a one-loop improvement over supersymmetric expectations.

\subsection{A fermionic model}
\label{sec:fermion}
	
	While somewhat more elaborate than the scalar model presented above, a fermionic model provides a clearer illustration of how the weak scale can be bounded by technically natural parameters in conjunction with the WGC. As in the scalar model, obtaining a non-trivial bound implies that charged fermions transforming under a new Abelian gauge group acquire some of their mass from electroweak symmetry breaking. To this end, we introduce a pair of vector-like fermions $L, N$ with conjugates $L^c, N^c$ transforming under both the SM and a new Abelian gauge group $U(1)_X$. The Weyl fermion $L$ has the SM quantum numbers $(1, {\bf 2})_{+\frac{1}{2}}$, while $N$ is a complete SM-singlet. For definiteness, we consider a $U(1)_X$ that gauges the sum of the $U(1)_V$ symmetries of the fermions, leading to charge assignments $Q[L] = Q[N] = -Q[L^c] = - Q[N^c] = +1$ under $U(1)_X$. 
	
	The following Dirac masses and couplings to the SM sector are consistent with existing symmetries:
	\begin{eqnarray}
	- \mathcal{L} \supset \left\{ m_L L L^c + m_N N N^c + y H^\dag L N^c + \tilde y H L^c N \right\} + {\rm h.c.}
	\end{eqnarray}
	After electroweak symmetry breaking, the corresponding mass terms read
	\begin{equation}
	- \mathcal{L} \supset \left( \begin{array}{cc} L & N \end{array} \right) \left( \begin{array}{cc} m_L & \frac{y v}{\sqrt{2}} \\ \frac{\tilde y v}{\sqrt{2}} & m_N \end{array} \right) \left( \begin{array}{c} L^c \\ N^c \end{array} \right) + {\rm h.c.} \ ,
	\label{eq:fermionmasses}
	\end{equation}
	where we have used $\langle H \rangle = v/\sqrt{2} \simeq 174$ GeV.
	The mass matrix contains a single physical phase, namely $\phi = {\rm arg}\left( y \tilde y m_L^* m_N^* \right)$.
	For concreteness, in what follows we focus on the case $\phi = \pi$, although the following discussion is largely independent of its exact value.
	
	At this point it is worth pausing to emphasize that the parameters $m_L, m_N, y$ and $\tilde y$ are all technically natural. In particular, $m_L, m_N \rightarrow 0$ respectively restore the axial symmetries $U(1)_{A,L}$ and $U(1)_{A,N}$, while $y, \tilde y \rightarrow 0$ restore the diagonal axial symmetries. Note that while $m_L$ and $m_N$ are formally independent parameters breaking the chiral symmetries of $L$ and $N$, since the Yukawa couplings only preserve the diagonal chiral symmetry this leads to radiative corrections that correlate the masses. In particular, we have
	\begin{equation}  \label{eq:loops}
		\delta m_N \approx \frac{y \tilde y}{4 \pi^2} m_L \log \frac{\Lambda_{\rm EFT}}{m_L} \ , \qquad {\rm and} \qquad \delta m_L \approx \frac{y \tilde y}{8 \pi^2} m_N \log \frac{\Lambda_{\rm EFT}}{m_N} \ ,
	\end{equation}
	which naturally correlates the two vector-like masses by a loop factor in theories without tuning.

	The physical mass spectrum following from Eq.(\ref{eq:fermionmasses}) consists of a single Dirac fermion carrying $U(1)_{\rm EM}$ charge with mass $m_L$, as well as two $U(1)_{\rm EM}$-neutral Dirac fermions, $\chi_1$ and $\chi_2$, with masses given by
	\begin{equation} \label{eq:chimasses}
		m_{\chi_{1,2}}^2 =  \frac{1}{2} \left( m_L^2 + m_N^2 + y^2 v^2 \mp (m_L - m_N) \sqrt{ (m_L + m_N)^2 + 2 y^2 v^2} \right) \ ,
	\end{equation}
	where we have set $|\tilde y| = |y|$ for convenience.
	All of the fermions carry unit charge under the unbroken $U(1)_X$. Holding $m_N, m_L$, and $y$ fixed, the mass of the lightest fermion increases as we increase the Higgs vev from $v = 0$.
	\footnote{This statement is not independent of the choice of $\phi$. For $\phi = 0$, $m_{\chi_1}$ initially decreases with $v$. However, both $m_{\chi_1}$ and $m_{\chi_2}$ ultimately increase as a function of $v$ for arbitrary values of the phase.}

	We can now ask what bounds on the parameter space might be implied by the electric WGC applied to $U(1)_X$. If the $U(1)_{\rm EM}$-charged fermion satisfies the electric WGC, then this would only imply $m_L \lesssim g M_{\rm Pl}$. In this case, the conjecture would hold for any value of $v$, and thus fails to give an interesting bound on the weak scale.
The more interesting possibility is for $m_L > m_N$ (subject to the correlation in Eq.(\ref{eq:loops})) so that for sufficiently small values of $v$ we have
	\begin{eqnarray} \label{eq:ewgcchi1}
	m_{\chi_1} \lesssim g M_{\rm Pl} \ , \qquad {\rm and} \qquad m_{\chi_\pm}, m_{\chi_2} \gtrsim g M_{\rm Pl} \ .
	\end{eqnarray}
	In this case the electric version of the conjecture can be violated as $v$ is increased, leading to a potentially interesting bound on the weak scale. Since $m_{\chi_1}$ depends on three technically natural parameters ($m_N, m_L, y$) and one UV-sensitive parameter ($v$), we can think of the WGC and fixed, technically natural values for $m_N, m_L, y$ and $g$ as setting an upper bound on $v$.
	From Eq.(\ref{eq:chimasses}), imposing $m_{\chi_1} \lesssim g M_{\rm Pl}$ leads to
	\begin{equation} \label{eq:fermionbound}
		v^2 \lesssim \frac{2}{y^2} \left( m_{\chi_1}^2 + m_{\chi_1} (m_L - m_N) - m_L m_N \right) \ .
	\end{equation}

As with the scalar model, we can compute the fine-tuning of the weak scale by treating $m_{\chi_1}$ (equivalently, $g$) as an input parameter fixed by the WGC. Assuming Eq.(\ref{eq:fermionbound}) is saturated, we have
		\begin{eqnarray}
	\Delta_{m_N} &=& \frac{2}{y^2 v^2} m_N (m_{\chi_1} + m_L) \ , \\
	\Delta_{m_L} &=& \frac{2}{y^2 v^2} m_L (m_{\chi_1} - m_N) \ , \\
	\Delta_{m_{\chi_1}} &=& \frac{2}{y^2 v^2} m_{\chi_1} (2 m_{\chi_1} + m_L - m_N) \ . \\
	\end{eqnarray}
	In order for the weak scale not to be overly sensitive to the underlying parameters ($\Delta \sim 1$), we typically require $y v \sim m_L \sim m_N$, implying that the fermions charged under $U(1)_X$ should be at or near the weak scale itself. We explore potential phenomenological implications of this model in section \ref{sec:pheno}.

\subsection{The magnetic WGC}
\label{sec:magneticWGC}
	
As the reader has no doubt been anticipating, we should also expect to satisfy the magnetic version of the conjecture. Whatever the physics implied by the magnetic WGC, the models in sections \ref{sec:scalar} and \ref{sec:fermion} already improve upon the $U(1)_{B-L}$ model in \cite{Cheung:2014vva} insofar as the magnetic scale can now lie at or above $v$, rather than well below it. That said, it warrants exploring what new physics might be implied by the magnetic WGC, and whether it modifies the presumed bound on the weak scale.

A straightforward way to address the magnetic version of the conjecture in the models presented here would be to embed the $U(1)_X$ into an $SU(2)_X$ gauge group, and to promote the scalar and fermion fields introduced in sections \ref{sec:scalar} and \ref{sec:fermion} to $SU(2)_X$ doublets. As discussed in the Introduction, a symmetry breaking pattern $SU(2)_X \rightarrow U(1)_X$ can be obtained though the non-zero vev of a Higgs field transforming in the adjoint representation.
The matter content charged under $U(1)_X$ would then be twice that of the minimal implementations discussed in previous sections. Moreover, if the scale of spontaneous symmetry breaking is close to $M_{\rm Pl}$, a $W$ gauge boson corresponding to the broken directions would be present in the spectrum at a scale $m_W \approx g_2 M_{\rm Pl} = 2 g M_{\rm Pl}$, and carrying 2 units of $U(1)_X$ charge. \footnote{$g_2$ refers to the $SU(2)_X$ gauge coupling, which corresponds to twice the charge quantum of the unbroken Abelian factor.} The phenomenological implications of such an extension will vary depending on the values of $g$, and the couplings between the $U(1)_X$-charged fields and the SM sector, as we discuss further in section \ref{sec:pheno}.

An important consideration concerning any implementation of the magnetic WGC is that it might alter our discussion leading to an upper bound on the weak scale. In particular, the $W$ gauge boson present in an $SU(2)_X$ embedding satisfies the electric version of the WGC itself, although with a higher upper bound on its mass, given that it carries twice the $U(1)_X$ charge of fundamental fields. Potentially, this could trivialize the electric WGC altogether, since the mass of $W$ is independent of $v$. A conclusive assessment of this potential caveat would require an answer to crucial questions regarding the form that the WGC takes in the far infrared: Can it be applied to a particle of arbitrary spin? Are there any circumstances under which the WGC particle is required to be the lightest state, or the state carrying the unit of charge? Does the super-extremal state need to be stable? (Notice that if the masses of both the $W$ gauge boson, and the $U(1)_X$-charged matter saturate their corresponding upper bounds, then the $W$ would be marginally unstable.) Any of these refinements, if true, would preserve the bound on the weak scale even in the presence of the $W$ gauge bosons.
These questions remain largely open, and answering them is perhaps one of the most urgent aspects of the WGC -- not only for the discussion presented here, but more generally regarding the standing of the conjecture in the context of the Swampland program \cite{Saraswat:2016eaz}.

As already mentioned in the Introduction, embedding into a non-Abelian gauge group is certainly not the only way of addressing the magnetic WGC. Alternative constructions could lead to their own phenomenological predictions, potentially quite different from the signs of naturalness that are common in standard solutions to the hierarchy problem. Another concrete possibility is the embedding of the $U(1)$ into the higher-dimensional spacetime symmetry group \cite{Wen:1985qj}. The fact that Kaluza-Klein theories contain magnetic monopole configurations was first noticed in \cite{Sorkin:1983ns,Gross:1983hb}, and in this case the scale of the magnetic WGC can be identified with the inverse size of the extra dimensions. Indeed, in the simplest case of an extra dimension compactified on $S^1$, we have $R^{-1} = g M_{\rm Pl}$ where $R$ is the radius of compactification and $g$ is the charge of the Kaluza-Klein $U(1)$. In this case, the magnetic WGC scale corresponds to the mass of the lightest KK modes. Now, one might naively expect that the appearance of an extra dimension at $g M_{\rm Pl} \sim v$ would cut off radiative corrections to the Higgs potential at the same scale, thereby replacing a non-trivial WGC bound on the weak scale with a more familiar solution in the form of large extra dimensions \cite{ArkaniHamed:1998rs}. But this is not the case: for a single extra dimension compactified on $S^1$, the five-dimensional Planck scale is $M_5 \sim (M_{\rm Pl}^2 R^{-1})^{1/3} \sim (v M_{\rm Pl}^2)^{1/3} \gg v$, rather than $M_5 \sim v$ as would be required to explain the weak scale purely by dilution of gravity in extra dimensions. The WGC applied to a $U(1)$ originating from the compactification of an extra dimension would appear to set a stronger bound on the weak scale than the extra dimension alone.

However, this appealing prospect is undone by the existence of the KK graviton modes \cite{Han:1998sg}, which trivialize the electric WGC bound. The massless five-dimensional graviton leads to four-dimensional KK modes with exactly $m = q = n/R$. Unlike the $W$ bosons in the non-Abelian completion, the $n=1$ KK graviton carries unit charge and would satisfy even more refined versions of the electric WGC. This exemplifies a message of \cite{Heidenreich:2015nta}, that gravity should `take care of itself' -- and, in this case, take care of everything!

Of course, these are but two specific options for physics associated with the magnetic WGC, and it would be valuable to explore a wider range of calculable possibilities. But these examples suffice to illustrate several key points regarding the role played by the magnetic version of the conjecture in any attempt to bound the weak scale. On one hand, these examples demonstrate that new physics appearing at the magnetic WGC scale $\Lambda$ need not be identified with the scale $\Lambda_{\rm EFT}$ characterizing UV contributions to the Higgs potential. On the other hand, they highlight the possibility that such new physics trivializes the bound on the weak scale coming from the electric WGC by introducing new super-extremal particles. Whether this occurs depends both on the specific realization of the magnetic conjecture and the precise form of its electric counterpart. Further exploration of these subtleties is crucial to determining the viability of possible WGC bounds on the weak scale.

\section{Scale separation \& the WGC with scalar fields}
\label{sec:scalarWGC}
	
	An alternative possibility is to separate the scale $\Lambda$ implied by the magnetic version of the conjecture from the bound on the super-extremal particle (and hence $v$) imposed by the electric WGC. This would accomplish two things. First, it would protect the electric bound on $v$ from being trivialized by whatever physics enters at $\Lambda$, since new particles appearing at $\Lambda$ would now be unlikely to satisfy the electric bound. Second, it would enable the magnetic WGC to be satisfied by a more violent modification of the theory (perhaps also cutting off the Higgs potential at the same scale) without implying additional degrees of freedom at the weak scale.

Such a separation is possible if the weak gravity conjecture with scalar fields (WGC+S) holds \cite{Palti:2017elp}, which in the simplest case implies that the mass of the WGC particle satisfies (taking for simplicity $q = 1$, and again ignoring overall factors of $\sqrt{2}$)
\begin{equation} \label{eq:swgc}
m^2 \lesssim (g^2 - \mu^2) M_{\rm Pl}^2 \ ,
\end{equation} 
where $\mu$ is the dimensionless coupling of the WGC particle to a light (ideally massless) scalar. The magnetic form of the WGC is unaltered, so if the WGC with scalar fields holds then the scales implied by the electric and magnetic forms of the WGC can be parametrically separated provided $g^2 - \mu^2 \rightarrow 0$ for finite $g, \mu$. As emphasized in \cite{Lust:2017wrl}, the separation between $m$ and $\Lambda$ implied by Eq.(\ref{eq:swgc}) suggests that the WGC can connect parametrically separated scales, potentially overriding expectations from naturalness. Even if the coincidence of $g$ and $\mu$ in Eq.(\ref{eq:swgc}) results purely from a tuning of parameters, it remains protected against radiative corrections.

Although explaining the coincidence between $g$ and $\mu$ was somewhat beside the point in \cite{Lust:2017wrl}, here we are interested in ensuring that the weak scale is not unduly sensitive to the choice of technically natural parameters involved in the bound. This motivates an underlying mechanism relating $g$ and $\mu$, since otherwise we would find $\Delta_{g,\mu} \propto \Lambda^2/v^2 \gg m^2/v^2$ in this scenario. There are several possibilities for generating the coincidence without fine-tuning, including approximate $\mathcal{N} = 2$ supersymmetry in the $U(1)_X$ sector or dynamics with a fixed point where $g_\star \approx \mu_\star$. However, whatever the underlying explanation -- including simple tuning of $g \approx \mu$ -- the parameters are technically natural and only run logarithmically, so the physics behind the coincidence can lie in the far UV.

In what follows, we will assume there is some mechanism that ensures $g \approx \mu$ in the UV, and explore the extent to which the WGC+S can be used to separate the electric and magnetic scales in the fermionic and scalar models presented in section \ref{sec:weakscale}. In this case, the weak scale will be fixed not by the requirement $m \lesssim g M_{\rm Pl}$, but by $m \lesssim \sqrt{g^2 - \mu^2} M_{\rm Pl}$. Insofar as $\sqrt{g^2 - \mu^2 } \ll g,$ this implies $\Lambda \sim g M_{\rm Pl}$ lies well above the weak scale and makes it unlikely for particles appearing at $\Lambda$ to satisfy the stronger bound. In addition to separating the mass $m$ of the WGC particle from the scale $\Lambda$, the WGC+S allows the weak scale to be stabilized by correspondingly larger values of $g$.  

A proof of principle for stabilizing the mass of a scalar using the WGC+S was presented in \cite{Lust:2017wrl}, including cases where the WGC particle is either the scalar itself or a fermion acquiring mass from the vev of the scalar. Neither of the examples presented in \cite{Lust:2017wrl} can be directly applied to the Higgs, but the necessary modifications are straightforward. As we will see, the additional fields required to connect a mass bound to the Higgs vev may limit the degree to which the WGC+S can be used to significantly separate scales, but this is still sufficient to protect against trivialization by the magnetic WGC.

\subsection{Retrofitting the scalar model}

To illustrate the potential to separate the electric and magnetic scales using the WGC+S, we first consider the simple scalar toy model of \cite{Lust:2017wrl}. This consists of a theory with gravity, a  $U(1)$ gauge group with coupling $g$, a charged scalar $h$ with charge $q = 1$ and mass $m_h$, and a neutral scalar $\varphi$ with mass $m_\varphi$. The scalar fields are coupled by a super-renormalizable coupling of the form
\begin{equation} \label{eq:swgcscalartoy}
- \mathcal{L} \supset 2 m_h \mu \varphi |h|^2
\end{equation}
where $\mu$ is the dimensionless strength of the scalar coupling in units of $m_h$, and $g \geq \mu$ is assumed. As $m_\varphi \rightarrow 0$, the WGC+S implies
\begin{equation} 
m_h^2 \lesssim  (g^2 - \mu^2) M_{\rm Pl}^2
\end{equation}
Several points bear emphasizing. First, the cubic coupling has been written in terms of $m_h$, which leads to scale separation for $g \approx \mu$. Ultimately this is just a choice of convention, as different choices may be absorbed into the definition of $\mu$, but it becomes relevant when attempting to explain the coincidence $g \approx \mu$ required for decoupling of the electric and magnetic scales. While expressing the cubic coupling in terms of $m_h$ is sensible when $h$ obtains all of its mass from the vev of $\varphi$, the scale of the cubic coupling is more broadly sensitive to the UV completion. For example, if the scale $m_h$ in Eq.(\ref{eq:swgcscalartoy}) is replaced by another scale $\rho$, the mass of the WGC particle must instead satisfy 
\begin{equation} 
m_h^2 \lesssim \left(g^2 - \mu_{\rm eff}^2 \right) M_{\rm Pl}^2 
\end{equation}
where $\mu_{\rm eff} = \mu \rho / m_h$. Separating this scale from the scale implied by the magnetic WGC would require a mechanism to generate the less obvious relation $g \approx \mu \rho / m_h$.

Second, although the WGC+S is strictly assumed to hold only when $\varphi$ is massless, the cubic coupling in Eq.(\ref{eq:swgcscalartoy}) generates a radiative correction of order $\delta m_\varphi \sim \mu m_h / 4 \pi$. Ultimately, this may vacate the scalar modification of the WGC bound: the scalar force now has finite range. Nonetheless, in \cite{Lust:2017wrl} it was argued that it should still be expected to hold provided $m_\varphi \ll m_h$, in which case $\varphi$ still gives rise to a long-range force on scales relevant to the WGC particle; this is of similar spirit to arguments in favor of the electric WGC in the presence of a small vector mass \cite{Reece:2018zvv}. For sufficiently small $m_\varphi$, the correction to the scalar-mediated force between two WGC particles at scales $\sim m_h^{-1}$  is of order
\begin{equation}
	\mu^2 \rightarrow \mu^2 \left( 1 - \mathcal{O} \left( \frac{m_\varphi}{m_h} \right) \right) \ ,
\end{equation}
and one might hope the WGC+S still holds provided $\mu^3 \ll g^2 - \mu^2 \ll \mu^2$. Of course, this parametric relation is again sensitive to the natural size of the cubic coupling. Similarly, if the scale $m_h$ in Eq.(\ref{eq:swgcscalartoy}) is replaced by another scale $\rho$, then preserving the WGC+S requires $\mu_{\rm eff}^3 \ll g^2 - \mu_{\rm eff}^2 \ll \mu_{\rm eff}^2$.

Third, even if a UV completion ensures $g \approx \mu$, it is necessary to ensure that this relationship is not spoiled by radiative corrections going into the infrared. As detailed in \cite{Lust:2017wrl}, this requires $g^4 \ll g^2 - \mu^2 \ll g^2$ and $\mu^4 \ll g^2 - \mu^2 \ll \mu^2$, both of which may be readily satisfied. 

While it is tempting to use the above toy model to stabilize the weak scale by identifying $h \rightarrow H$, this would require $H$ to be charged under the $U(1)$ gauge group. Preserving a WGC bound on $H$ consistent with electroweak symmetry breaking and couplings to SM fermions is quite challenging, and instead motivates retrofitting the scalar model presented in section \ref{sec:scalar}. In addition to the scalar $\Phi$ charged under a $U(1)_X$ gauge group, we add a light scalar $\varphi$ with a super-renormalizable coupling to $\Phi$:
\begin{equation}
	- \mathcal{L} \supset m^2 |\Phi|^2 + \lambda |\Phi|^4 + \kappa |\Phi|^2 |H|^2 + 2 \mu \rho \varphi |\Phi|^2 \, .
\label{eq:scalarswgc}
\end{equation}
Here we have written the $\varphi$-$\Phi$ coupling in terms of a dimensionless parameter $\mu$ and a dimensionful scale $\rho$, whose size we do not assume in advance. Radiative corrections will also generate a coupling between $\varphi$ and $|H|^2$, but this is naturally small enough to avoid interfering with the mechanism.

The WGC with scalars in this case requires that $\Phi$ satisfy the bound
\begin{equation} \label{eq:scalarswgcbound}
m_\Phi^2 \lesssim \left(g^2 - \mu^2 \frac{\rho^2}{m_\Phi^2} \right) M_{\rm Pl}^2 \ ,
\end{equation}
and thus one might hope to separate the electric and magnetic scales with a mechanism that ensures $g^2 \approx \mu^2 \rho^2/m_\Phi^2$. However, it bears emphasizing that the additional contribution to $m_\Phi$ from electroweak symmetry breaking makes understanding the separation of electric and magnetic scales in this case more challenging than in the toy model of \cite{Lust:2017wrl}. Even if the UV completion sets $\rho = m$ and $\mu = g$ (perhaps the most one could hope for in the UV from an underlying mechanism), the contribution to $m_\Phi$ from the Higgs vev makes it difficult to send $g^2 - \mu^2 \rho^2 / m_\Phi^2 \rightarrow 0$ without rendering the weak scale sensitive to underlying parameters at the level of $\Delta \sim m_\Phi^2 / \kappa v^2 \gg 1$. Although significant scale separation does not seem possible without tuning, separation of any amount is radiatively stable and could be sufficient to prevent particles associated with the magnetic WGC from trivializing the bound on the weak scale.

\subsection{Retrofitting the fermionic model}

A fermionic toy model was also presented in \cite{Lust:2017wrl}, consisting of a theory with gravity, a  $U(1)$ gauge group with coupling $g$, a charged Dirac fermion $\psi$ with charge $q = 1$ and mass $m_\psi$, and a neutral scalar $\varphi$ with mass $m_\varphi$ and Yukawa coupling
\begin{equation} \label{eq:swgcfermiontoy}
- \mathcal{L} \supset  \mu \varphi \psi \bar \psi
\end{equation}
In this case there is no ambiguity regarding the strength of the Yukawa coupling, and the WGC+S requires $m_\psi^2 \lesssim (g^2 - \mu^2) M_{\rm Pl}^2$. To bound a scalar mass in this example, \cite{Lust:2017wrl} then set $m_\psi = 0$ and introduce a neutral scalar $h$ of mass $m_h$ coupling to the Dirac fermion via 
\begin{equation} \label{eq:swgcfermiontoy2}
- \mathcal{L} \supset  \mu_h h \psi \bar \psi
\end{equation}
The mass of the fermion arises entirely from the vacuum expectation value of $h$, bounding the scalar mass. Preserving $g \approx \mu$ requires $g^4 \ll g^2 - \mu^2 \ll g^2$ and $\mu^4 \ll g^2 - \mu^2 \ll \mu^2$ as in the scalar toy model, as well as $\mu^2 \mu_h^4 \ll g^2 - \mu^2 \ll \mu^2$. Although not noted in \cite{Lust:2017wrl}, radiative corrections induce a mass for $\varphi$ at one loop of the form
\begin{equation}
	\delta m_\varphi^2 \sim \frac{\mu^2 \Lambda_{\rm EFT}^2}{16 \pi^2} \ , 
\end{equation}
where we assume $\Lambda_{\rm EFT}$ can also be identified with the cutoff of the $\varphi$ potential. In contrast to the scalar toy model, this mass correction is UV sensitive. The ultimate sign of $m_\varphi^2$ is unimportant, as $\langle \varphi \rangle \ll m_\psi$ provided $\Lambda_{\rm EFT} \ll M_{\rm Pl}$, and the WGC+S should still apply to the radial mode of $\varphi$ if it acquires a vev. Even so, ensuring that the Yukawa force is not significantly modified at separations of order $m_\psi^{-1}$ requires
\begin{equation}
\frac{\mu \Lambda_{\rm EFT}}{4 \pi} \ll m_\psi \ .
\end{equation}
Given that $\mu \ll 1$, the above inequality can be satisfied if $\Lambda_{\rm EFT} \sim \Lambda$.
Finally, radiative corrections induce a linear mixing between $\varphi$ and $h$, but this can be rendered innocuous if $\mu_h$ is sufficiently small.

As before, this toy model cannot be applied directly to the weak scale, in this case due to the gauge quantum numbers of $H$, but again it may be suitably generalized to the models in section \ref{sec:weakscale}. One might initially be tempted to retrofit the $U(1)_{B-L}$ model of section \ref{sec:noBminusL} to leverage the WGC+S, but this is a non-starter. While the WGC with scalars could be used to separate neutrino masses from the magnetic WGC scale, experimental bounds on the $U(1)_{B-L}$ coupling still force $g M_{\rm Pl} \lesssim 1$ keV, and all the problems discussed in section \ref{sec:noBminusL} remain. A related option would be to gauge other symmetries of the SM with weaker constraints, such as non-universal family symmetries, which have received increased attention in recent years in light of flavor anomalies (see e.g.~\cite{Crivellin:2015lwa}). For example, one could consider gauging $L_\mu - L_\tau$ number \cite{Foot:1990mn,He:1990pn}, under which the lightest SM neutrino is the lightest charged state. Much weaker bounds on the gauge coupling in this case mean that the WGC with scalars could be used to raise the magnetic WGC scale to the weak scale or above while keeping neutrino masses fixed around an eV. However, in all of these scenarios, the force-carrier $\varphi$ must be charged under either the SM or the new gauge symmetry. The former option is highly constrained, while for the latter option the force-carrier is itself a light super-extremal particle (albeit one carrying twice the charge of the neutrinos, making the fate of the weak scale potentially subject to the same subtleties discussed in section \ref{sec:magneticWGC}).

More promisingly, the fermionic model presented in section \ref{sec:fermion} can be simply retrofitted to leverage the WGC+S. In addition to the vector-like fermions $L, L^c$, $N, N^c$ charged under a $U(1)_X$ gauge group, we add a light scalar $\varphi$ with appropriate Yukawa couplings:
\begin{eqnarray}\label{eq:fermionswgc}
	- \mathcal{L} \supset \left\{ (m_L + \tilde \mu \varphi) L L^c + (m_N + \mu \varphi) N N^c + y H^\dag L N^c + \tilde y H L^c N \right\}+  {\rm h.c.}
\end{eqnarray}

The various couplings in Eq.(\ref{eq:fermionswgc}) give rise to radiative corrections that might imperil $g \approx \mu$ and $m_\varphi^2 \sim 0$. At one loop, the $U(1)_X$ coupling runs proportional to itself, $\delta g \sim g^3 / 16 \pi^2$, which is innocuous provided $g^2 \lesssim \sqrt{g^2 - \mu^2} \ll g$ -- clearly the case for the infinitesimal values of $g$ of interest. At the same loop order, the Yukawa coupling $\mu$ accumulates various contributions:
\begin{enumerate}
\item Running proportional to itself, $\delta \mu \sim \mu^3 / 16 \pi^2$, which is innocuous provided $\mu^2 \lesssim \sqrt{g^2 - \mu^2} \ll \mu$.
\item Running proportional to $\mu$ and $g^2$, $\delta \mu \sim \mu g^2 / 16 \pi^2$, which is innocuous provided $g^2 \lesssim \sqrt{g^2 - \mu^2} \ll g$.
\item Running proportional to $\tilde \mu$ and $y, \tilde y$, $\delta \mu \sim \tilde \mu y \tilde y / 16 \pi^2,$ which is innocuous provided $y \tilde y \tilde \mu / \mu \lesssim \sqrt{g^2 - \mu^2} \ll \mu$. 
\end{enumerate} 
While the first two conditions are easily satisfied, this last requirement is in some tension with the preference for large $y, \tilde y$ discussed in section \ref{sec:fermion} but may be ameliorated by taking $\tilde \mu$ small. Of course, $\tilde \mu$ may not be made arbitrarily small without tuning since it is generated at one loop proportional to $y \tilde y \mu$, but this hierachy alone is sufficient to avoid spoiling $g \approx \mu$.

In addition, radiative corrections induce a mass for $\varphi$ at one loop of the form
\begin{eqnarray}
	\delta m_{\varphi}^2 = - \left(2 \tilde \mu^2 + \mu^2 \right)\frac{\Lambda_{\rm EFT}^2}{8 \pi^2} \ .
\end{eqnarray}
Insofar as $\mu, \tilde \mu \sim g \ll 1$, we can easily preserve $m_\varphi \ll m_{\chi_1}$ if, e.g.~$\Lambda_{\rm EFT} \sim \Lambda$. In this case, assuming the inequalities required to preserve $g \sim \mu$ are satisfied and $m_\varphi$ is of the same order as its radiative corrections, we have 
\begin{eqnarray}
m_{\varphi} \ll m_{\chi_1} \ll \Lambda \, .
\end{eqnarray}

However, as in the scalar example, there is an important complication arising from the contribution to the fermion masses from electroweak symmetry breaking. After electroweak symmetry breaking, the Yukawa coupling of the lightest fermion mass eigenstate to $\varphi$ depends on mixing angles. If a dynamical mechanism sets $g = \mu$ in the UV, this could be spoiled by mixing effects in the IR that imbue the lightest fermion mass eigenstate with a significantly different Yukawa coupling. Perhaps the best-case scenario corresponds to the limit $|m_L - m_N| \gg y v$, in which case the lightest mass eigenstate is primarily composed of $N, N^c$. The mixing angle in this limit is proportional to $|y v| / m_L$, and for $\Delta \sim 1$ this corresponds to a minimum fractional shift of order $\sqrt{m_N/m_L} \sim y/4 \pi$. Thus if a UV mechanism sets $\mu = g$, the fractional shift in the Yukawa coupling to the lightest mass eigenstate suggests that the WGC+S can separate $m_{\chi_1}$ and $\Lambda$ without tuning in this example by
\begin{equation}
\frac{m_{\chi_1}}{\Lambda} \gtrsim \frac{y}{4 \pi} \, .
\end{equation}
While modest, this separation is can still suffice to protect the bound on the weak scale from being trivialized by states associated with the magnetic WGC.

\section{Phenomenology}
\label{sec:pheno}

As we have seen, although the WGC appears to circumvent EFT expectations, its application to the hierarchy problem nonetheless shares features common to more conventional approaches: new particles at or below the weak scale that couple to the Higgs. This leads to fairly sharp experimental targets for probing the WGC's role in bounding the Higgs mass. In both the scalar and fermionic models introduced in section \ref{sec:weakscale}, the dominant signatures are those of the Higgs portal, leading to constraints from invisible Higgs decays for $m_{\Phi}$, $m_{\chi_1} < m_h / 2$ and radiative corrections over a wider range of masses. 

The application of the WGC to the weak scale shares another feature of more conventional approaches to the hierarchy problem: a stabilization symmetry (in this case, the new $U(1)$ itself) that furnishes potential DM candidates. As with more conventional approaches such as supersymmetry, the motivated parameter space of these models can be further shaped by requiring viable DM. Since the lightest particle charged under an unbroken gauge symmetry is guaranteed to be stable, the most attractive possibility is that the super-extremal state itself accounts for the observed DM relic density.
Although the details of this ``super-extremal DM'' scenario depend on the specific model under consideration, we identify some universal features. Specifically, if the DM particle saturates the WGC bound, then the DM behaves as a collisionless plasma, with collective effects dominating over nearest-neighbour interactions. On very large scales, plasma instabilities can lead to dissipative dynamics, with potentially observable consequences.

In section \ref{sec:higgsportal} we explore the Higgs portal signatures of the scalar and fermion models introduced in \ref{sec:scalar} and \ref{sec:fermion}, respectively, while in sections \ref{sec:scalarpheno} and \ref{sec:fermionpheno} we discuss the details of DM that approximately saturates the WGC bound in the context of each model. In \ref{sec:plasma} we estimate the timescales and distances over which the DM plasma effectively behaves as collisional if instabilities can grow, and show that such collective effects may affect the DM distribution in the context of galaxy cluster collisions.

\subsection{Collider signals}
\label{sec:higgsportal}

Given that connecting weak gravity to the weak scale seems to require new (light) particles coupling to the Higgs, signatures associated with the Higgs provide the most direct avenues for discovery at colliders. Although cosmological considerations may further shape the viable parameter space, they may be obviated by competing effects in the early universe (e.g.~low reheating temperatures, late decays, or other phenomena altering standard expectations for relic abundances). As such, we will explore the collider phenomenology without reference to potential cosmological constraints, reserving those for subsequent consideration in sections \ref{sec:scalarpheno} and \ref{sec:fermionpheno}.

\subsubsection*{Scalar model}

\begin{figure}
   \centering
      \includegraphics[scale=0.8]{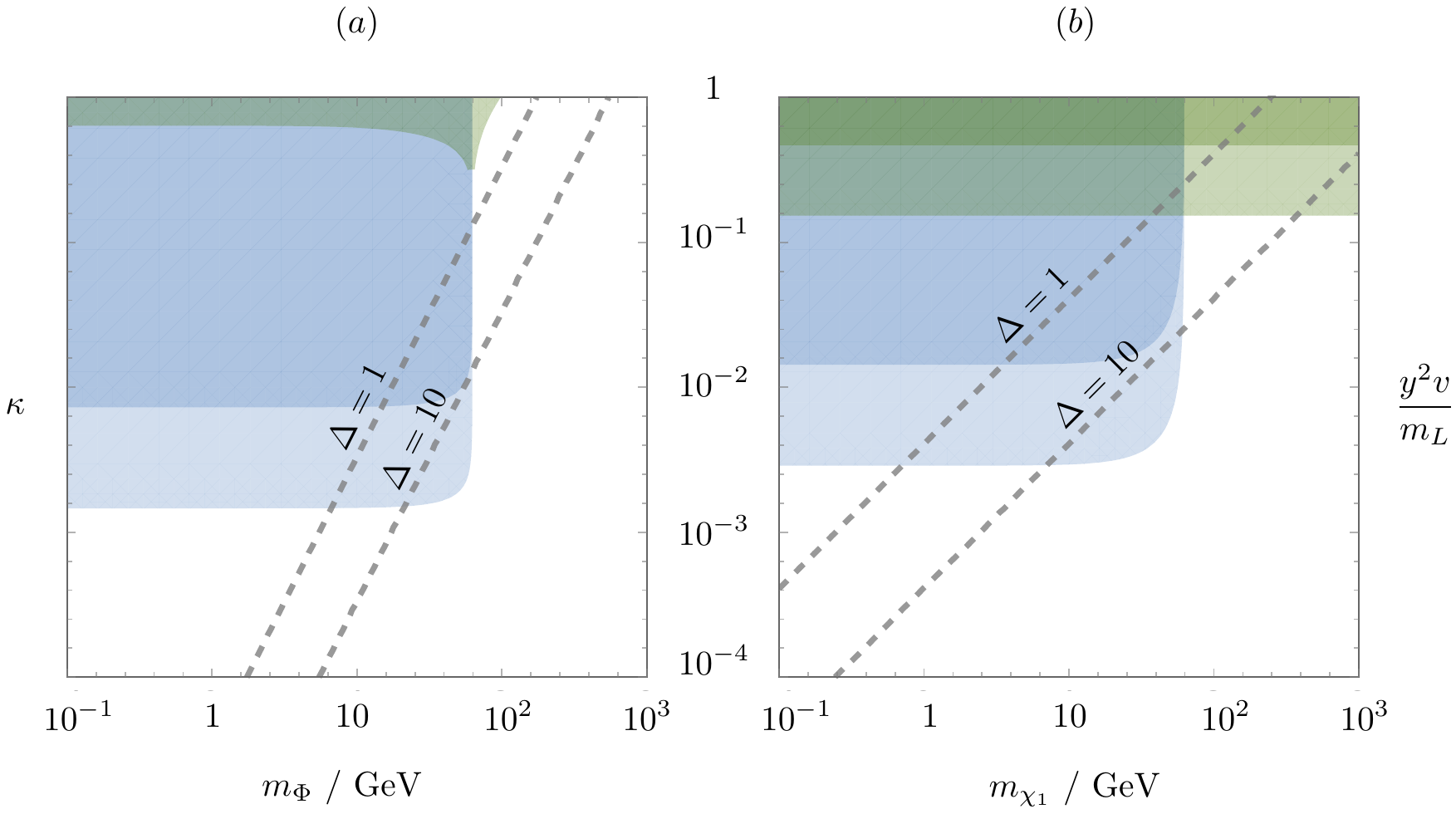}
   \caption{{\bf (a)} Constraints on the parameter space of the scalar model as a function of $m_\Phi$ and $\kappa$ coming from the current upper bound ${\rm BR(} h \rightarrow {\rm invisible )} < 20\%$ (dark blue), as well as the ${\rm BR(} h \rightarrow {\rm invisible )} < 1\%$ region that will be accessible in the near future (light blue). The light green region denotes the parameter space accessible by a percent-level measurement of the $hZ$ cross section achievable at future lepton colliders. The dashed lines correspond to $\Delta = 1, 10$. {\bf (b)} Constraints on the parameter space of the fermionic model as a function of $m_{\chi_1}$ and $y^2 v / m_L$ in the limit $m_N, v \ll m_L$ coming from the current upper bound ${\rm BR(} h \rightarrow {\rm invisible )} < 20\%$ (dark blue), as well as the ${\rm BR(} h \rightarrow {\rm invisible )} < 1\%$ region that will be accessible in the near future (light blue). The dark green region denotes the $2 \sigma$ exclusion from current bounds on the $T$ parameter, while the light green region denotes the potential exclusion from bounds on the $T$ parameter achievable at future lepton colliders. The dashed lines correspond to $\Delta = 1, 10$.}
 \label{fig:pheno}
\end{figure}

The primary signatures of the scalar model presented in section \ref{sec:scalar} are those of the $\mathbb{Z}_2$-symmetric Higgs portal \cite{Silveira:1985rk, McDonald:1993ex, Burgess:2000yq}. At present, Higgs decays $h \rightarrow \Phi \Phi^*$ (when kinematically accessible) provide the most promising collider probe of this scenario, though radiative corrections to Higgs properties and off-shell production of $\Phi \Phi^*$ via the Higgs portal are likely to be relevant at proposed future colliders. The left panel of Figure~\ref{fig:pheno} shows the region of parameter space in terms of $\kappa$ and $m_\Phi$ that is ruled out by the current upper bound of $20\%$ on the invisible branching fraction of the Higgs \cite{Khachatryan:2016whc,Aad:2015pla,Aaboud:2017bja,Aad:2015txa,Sirunyan:2018owy,Aaboud:2018sfi}. 
As indicated, attainment of $1\%$ sensitivity in the future \cite{CMS:2017cwx} would significantly extend coverage. 

For $m_\Phi > m_h/2$, $\Phi$ may be pair-produced via an off-shell Higgs, though prospects for discovery in this channel at the LHC and future colliders are fairly modest \cite{Craig:2014lda}. Perhaps the most promising avenue for probing $m_\Phi > m_h/2$ is via radiative corrections to Higgs properties, which can be most strongly constrained by a precision measurement of the $e^+ e^- \rightarrow hZ$ cross section at future lepton colliders \cite{Englert:2013tya, Craig:2013xia}. The potential reach of a $1\%$ measurement of the fractional shift $\delta \sigma_{hZ}$ is also shown in the left panel of Figure~\ref{fig:pheno}. The combination of percent-level sensitivity to both Higgs invisible decays and the $hZ$ cross section would cover much of the motivated parameter space for $m_\Phi \gtrsim 5$ GeV.

\subsubsection*{Fermionic model}

Collider signatures of the fermionic model presented in section \ref{sec:fermion} are primarily those of Dirac singlet-doublet DM \cite{Freitas:2015hsa,Fedderke:2015txa, Yaguna:2015mva}. The most promising signatures ultimately depend on the value of the doublet mass parameter $m_L$, as direct searches at colliders are sensitive to new vector-like doublet fermions below the TeV scale. While the full range of phenomenology depends in detail on the underlying parameters $m_N, m_L$, and $y$, for the sake of illustration we will focus here on the limit $v, m_N \ll m_L$. In this limit, the heavy doublet fermion can be integrated out, leaving a singlet-like Dirac fermion coupled to the Higgs. The phenomenology can be characterized in terms of just two parameters ($m_N$ and $m_L / y^2$), while the dominant constraints come from precision electroweak bounds and Higgs invisible decays.

The strongest constraint comes from invisible Higgs decays. The right panel of Figure \ref{fig:pheno} shows the region of parameter space (in the limit $v, m_N \ll m_L$) in terms of $m_{\chi_1}$ and the dimensionless combination $y^2 v / m_L$ that is ruled out by the current upper bound of $20\%$ on the invisible branching fraction of the Higgs, as well as the region that would be probed by a percent-level measurement of the same channel.

Prospects for discovering pair production of $\chi_1$ via an off-shell Higgs for $m_{\chi_1} > m_h/2$ are no better than in the scalar model. While radiative corrections to Higgs couplings from the light singlet-like fermion are promising, a stronger limit arises from precision electroweak constraints \cite{Fedderke:2015txa}. For large $y$ where precision electroweak constraints are relevant, contributions to $T$ are enhanced over contributions to $S$ by a relative factor of $y^2/e^2$, and so we focus on the former for simplicity. Current limits on the $T$ parameter in the limit $v, m_N \ll m_L$ are shown in the right panel of Figure \ref{fig:pheno}, as well as the potential reach of improved precision electroweak measurements at proposed lepton colliders \cite{Fedderke:2015txa}. Much like in the scalar case, the combination of percent-level sensitivity to Higgs invisible decays and improved precision electroweak measurements attainable at a future lepton collider would cover much of the motivated parameter space for $m_{\chi_1} \gtrsim 1$ GeV at large $m_L$. Of course, for smaller values of $m_L$, properties of the additional fermion states can provide more spectacular signatures, including direct electroweak production of charged and neutral fermions and loop-level corrections to $h \rightarrow \gamma \gamma$.

\subsection{Scalar dark matter}
\label{sec:scalarpheno}

Given that the lightest particle charged under $U(1)_X$ is stable, the super-extremal particle in both the scalar and fermionic models is a natural DM candidate, and imposing astrophysical constraints further shapes the motivated parameter space. In the scalar model, the super-extremal particle couples exclusively to the SM through the $\mathbb{Z}_2$-symmetric Higgs portal, and thus would correspond to Higgs portal DM where the DM is charged under an additional Abelian gauge group. Moreover, since we will be interested in values of the $U(1)_X$ gauge coupling $g \lll 1$, the quartic coupling $\kappa$ is the only interaction relevant for DM production. Depending on its size, the observed relic abundance may be obtained dynamically either through freeze-out or freeze-in.
The former possibility, however, appears to be in tension with the latest direct detection constraints from XENON1T \cite{Aprile:2018dbl}, as perturbative values of $\kappa$ leading to the correct DM density result in too large a scattering cross section with nuclei (see \cite{Arcadi:2019lka} for an extensive review).

We focus instead on the alternative possibility that the DM is produced through freeze-in \cite{Hall:2009bx}.
The possibility of scalar DM that is a SM singlet, and whose relic density is produced via Higgs portal freeze-in was first discussed in \cite{Yaguna:2011qn,Frigerio:2011in,Chu:2011be}.
In the regime $m_\Phi \ll m_h / 2$ the dominant DM production channel is through Higgs decays $h \rightarrow \Phi \Phi^*$, and the DM abundance is given by
 \begin{equation}
	\Omega_\Phi h^2 \sim 0.1 \left( \frac{\kappa}{3 \cdot 10^{-12}} \right)^2 \left( \frac{m_\Phi}{1 \ {\rm GeV}} \right) \left( \frac{246 \ {\rm GeV}}{v} \right) \ .
 \end{equation}
The value of $\kappa$ leading to the observed DM relic density may thus be written as
\begin{equation}
	\kappa \approx 3 \cdot 10^{-12} \left( \frac{1 \ {\rm GeV}}{m_\Phi} \right)^{1/2} \left( \frac{v}{246 \ {\rm GeV}} \right)^{1/2} \ ,
	\label{eq:kappaDM}
\end{equation}
with DM being overproduced for larger couplings.
Notice that in this regime the DM is relativistic when produced, and therefore $m_\Phi \gtrsim 1 \ {\rm keV}$ is required for the DM to be cold.

As discussed in section \ref{sec:scalar}, in the absence of fine-tuning one would have $m_\Phi \sim \sqrt{\kappa} v$, and the DM mass is set only by the connector coupling and the value of the weak scale.
In this case, the dependence on $v$ in Eq.(\ref{eq:kappaDM}) drops out, and the correct relic density is achieved for
\begin{equation}
	\kappa \approx 8 \cdot 10^{-11} \qquad \qquad \left( {\rm when} \ m_\Phi = \sqrt{\kappa / 2} v \ \right) \ .
	\label{eq:kappaDMnatural}
\end{equation}

\begin{figure}
   \centering
   \includegraphics[scale=0.8]{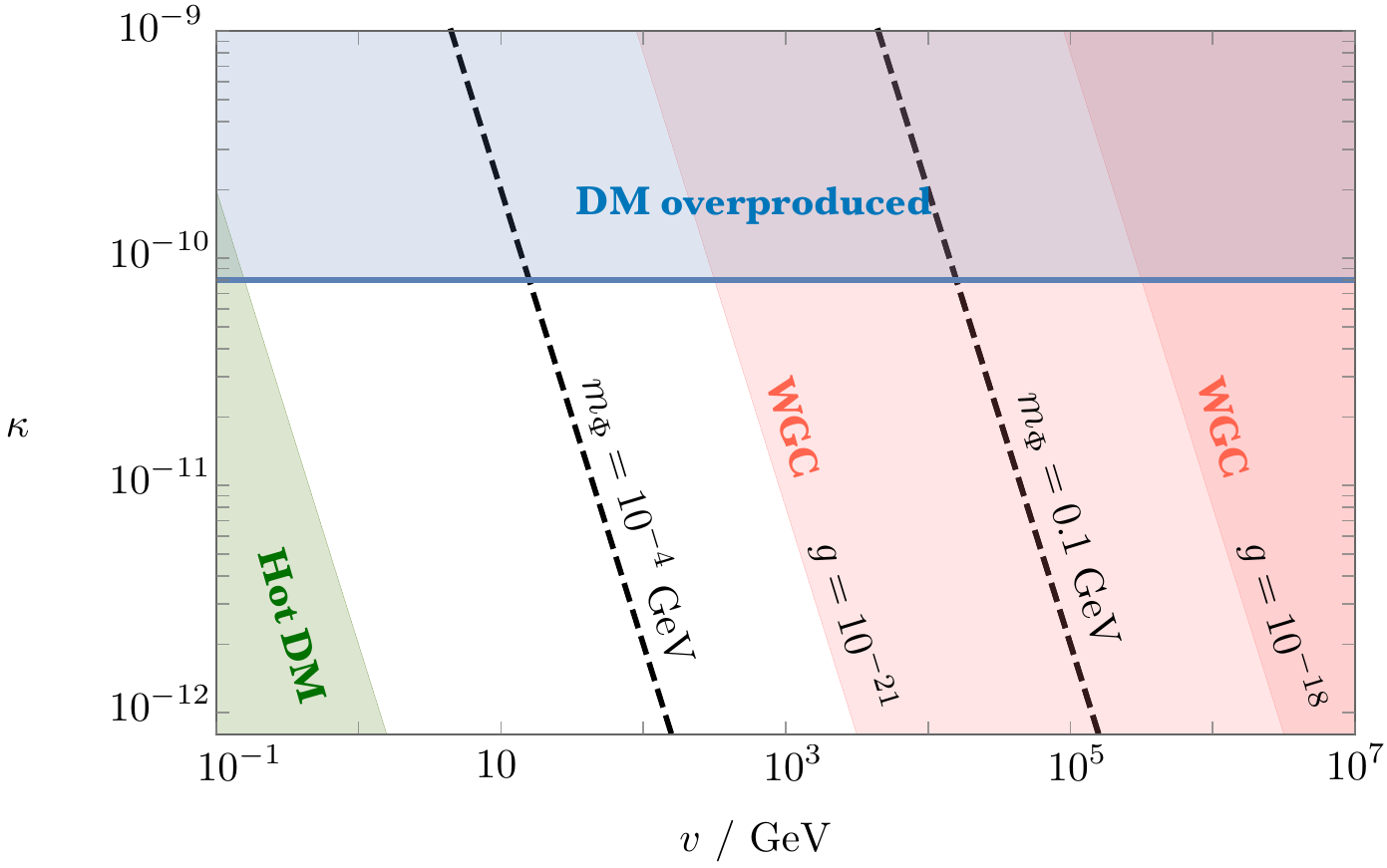} 
   \caption{Everywhere in this plot $m_\Phi = \sqrt{\kappa / 2} v$.
   The correct DM relic density is obtained through Higgs portal freeze-in along the horizontal blue line, corresponding to the value of $\kappa$ in Eq.(\ref{eq:kappaDMnatural}), with the region above the line corresponding to DM overproduction.
   The green region corresponds to $m_\Phi \lesssim 1 \ {\rm keV}$, and is ruled out as DM would be hot.
   The pink region corresponds to $\Phi$ being sub-extremal for the values of the gauge coupling $g$ indicated in the figure, and therefore is ruled out by the WGC.}
   \label{fig:scalarDM_v}
\end{figure}

Figure~\ref{fig:scalarDM_v} shows the region of parameter space in terms of the connector coupling $\kappa$, and the value of $v$, in which $\Phi$ can successfully account for the observed DM abundance, under the assumption that $m_\Phi \approx \sqrt{\kappa / 2} v$. If $g \sim 10^{-21}$, the WGC sets an upper bound on the DM mass $m_\Phi \lesssim 10^{-3} \ {\rm GeV}$, which in turn constrains the value of the weak scale to be below $\sim 1 \ {\rm TeV}$.
On the other hand, the requirement that the DM must be cold imposes a lower bound $m_\Phi \gtrsim 1 \ {\rm keV}$, which in turn implies $v \gtrsim 1 \ {\rm GeV}$. In this context, this lower bound may be regarded as an anthropic requirement on the minimal value of the weak scale, as smaller values of $v$ would lead to the DM being hot, precluding structure formation to proceed as observed.

\begin{figure}
   \centering
   \includegraphics[scale=0.8]{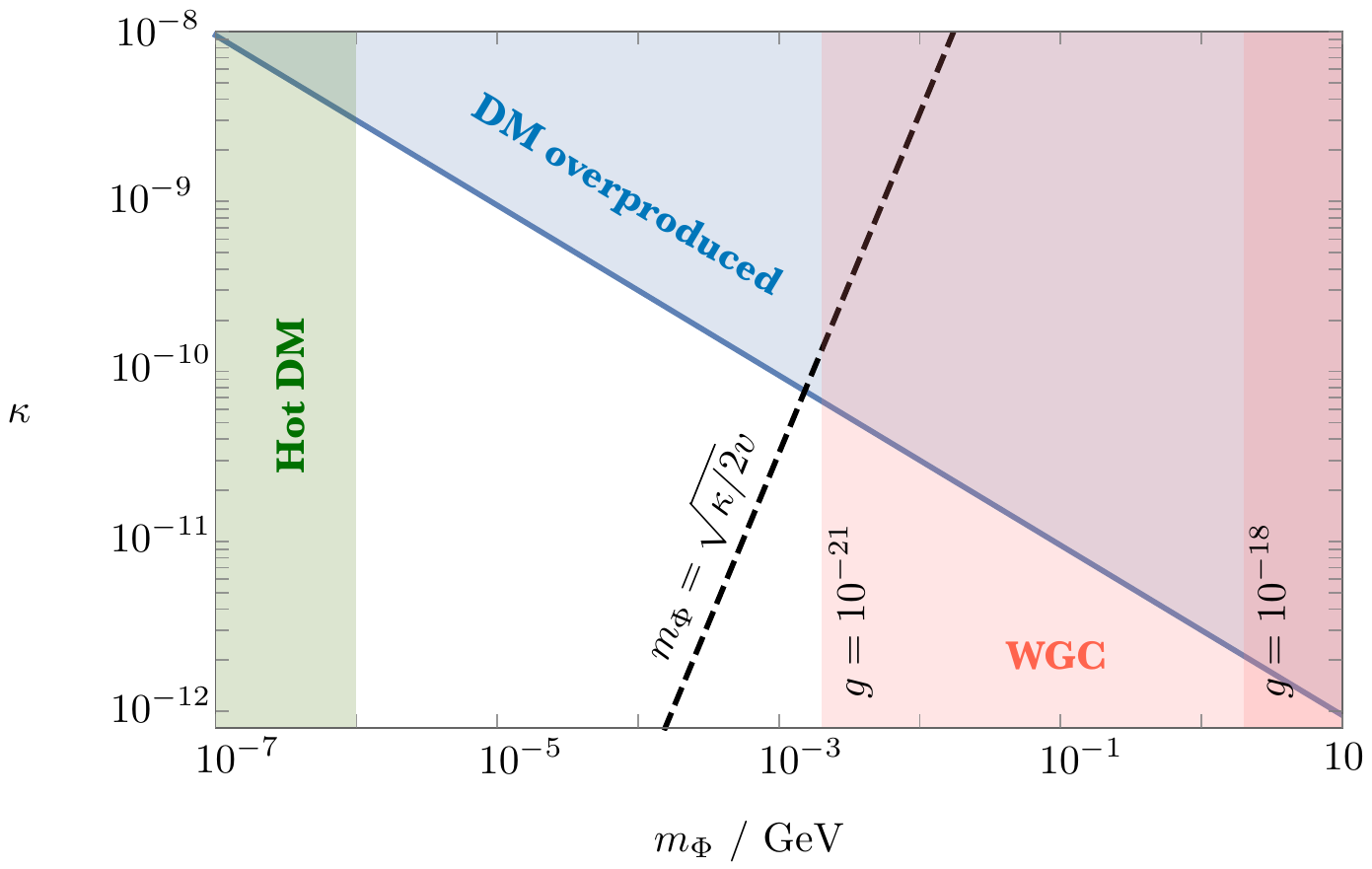} 
   \caption{Everywhere in this plot $v = 246 \ {\rm GeV}$.
   The correct DM relic density is obtained through Higgs portal freeze-in along the horizontal blue line, as given Eq.(\ref{eq:kappaDM}), with the region above the line corresponding to DM overproduction.
   The green region corresponds to $m_\Phi \lesssim 1 \ {\rm keV}$, and is ruled out as DM would be hot.
   The pink region corresponds to $\Phi$ being sub-extremal for the values of the gauge coupling $g$ indicated in the figure, and therefore is ruled out by the WGC.
   The dashed line corresponds to the case $m_\Phi = \sqrt{\kappa / 2} v$, and can therefore be considered natural (see discussion in the text), while the region to the left of this line would require fine-tuning.}
   \label{fig:scalarDM_mDM}
\end{figure}

Alternatively, one can also look at the viable region of parameter space in terms of $\kappa$ and the mass of the DM particle, while keeping $v$ fixed to its observed value. This is shown in Figure~\ref{fig:scalarDM_mDM}, where the dashed line corresponds to the case $m_\Phi = \sqrt{\kappa / 2} v$, and results in the observed relic density for $m_\Phi \approx 10^{-3} \ {\rm GeV}$.


We can further address the magnetic version of the conjecture by, for example, embedding the Abelian factor into an $SU(2)_X$ gauge group.
As already mentioned in section \ref{sec:scalar}, $\Phi$ would then be promoted into an $SU(2)_X$ doublet, with components $\Phi_1$ and $\Phi_2$ coupling to the Higgs with the same quartic coupling $\kappa$.
Assuming $m_{\Phi_1} \approx m_{\Phi_2} \approx \sqrt{\kappa / 2} v$, the only difference with respect to the situation discussed above is that the value of $\kappa$ required to obtain the observed relic abundance would now be smaller by a factor of $\sqrt{2}$, in order to account for the doubling of degrees of freedom in the DM sector.
An extra vector degree of freedom, the $W$ gauge boson associated to the broken directions, would also be part of the spectrum, making an appearance below the scale $\sim g M_{\rm Pl}$, which parametrically coincides with the DM mass if the electric WGC bound is saturated. However, the values of $g$ that we consider here are so exceedingly small that interactions between this massive vector and the components of $\Phi$ would have no implications regarding DM production.

In fact, given that both $g, \kappa \lll 1$, one might fear that this model has no experimental implications.
Indeed, the values of $\kappa$ required to produce DM through freeze-in via the Higgs portal are so small that it is probably impossible to directly probe this production mechanism \cite{Chu:2011be}.
There is some hope, however, that the DM being a plasma of $\Phi$ and $\Phi^*$ particles subject to long-range interactions may have observable consequences, even for the tiny values of the gauge coupling considered here.
We defer the discussion of potential collective effects in the DM plasma to section \ref{sec:plasma}, as these would not be unique to the specific scalar model discussed here, but rather common to any model of plasma DM where the DM mass is not too far below the super-extremality bound.

\subsection{Fermionic dark matter}
\label{sec:fermionpheno}

The fermion model of section \ref{sec:fermion} also admits a DM interpretation, with the lightest fermion charged under the $U(1)_X$ gauge group, $\chi_1$, playing the role of the DM particle.
Unlike the situation for the scalar case discussed in \ref{sec:scalarpheno}, the additional field content introduced in the fermion model is charged under the SM gauge group, and, as a result, the viable region of parameter space is subject to significant experimental constraints, both from DM direct detection as well as collider experiments.
Models of DM based on the field content of Eq.(\ref{eq:fermionmasses}) have been thoroughly studied in the literature (see e.g.~\cite{Freitas:2015hsa}), so we will limit ourselves here to a summary of those aspects that are relevant for our discussion. 

As discussed in \cite{Freitas:2015hsa}, a DM abundance of $\chi_1$ particles can be obtained through the standard freeze-out mechanism. However, direct detection constraints require the DM to be dominantly an electroweak singlet. Moreover, the correct relic density is obtained only in the region of parameter space where the mass difference between the two EM neutral fermions, $\chi_1$ and $\chi_2$, is of order a $\sim {\rm few} \ \%$, so that co-annihilation is active.
This can be achieved, in the notation of section \ref{sec:fermion}, for $m_N, m_L \ll m_L - m_N \ll y v$. 
For a value of the physical phase $\phi = \pi$ (defined in \ref{sec:fermion}), one finds
\begin{equation}
	m_{\chi_{1 (2)}} \approx m_{N (L)} + \frac{(y v)^2}{2 (m_L + m_N)} \ ,
\label{eq:fermionDMmass}
\end{equation}
so that
\begin{equation}
	\delta \equiv \frac{m_{\chi_2} - m_{\chi_1}}{m_{\chi_1}} \approx \frac{m_L - m_N}{m_N} \ .
\end{equation}

One can then choose $m_L, m_N \sim 100 \ {\rm GeV}$, and set $\delta \sim 0.01$ by choosing $m_L - m_N$ small -- a technically natural fine-tuning.
Both $\chi_1$ and $\chi_2$ would thus appear roughly at the weak scale, separated by a small mass difference.
As an example, values of the vector-like masses given by $m_N \simeq 295 \ {\rm GeV}$ and $m_L \simeq 305 \ {\rm GeV}$, as well as a Yukawa coupling $y \simeq 5 \cdot 10^{-4}$, leading to $\delta \simeq 3 \%$, can result in the correct relic density, and are consistent with current experimental constraints.
 
An upper bound on the mass of the DM particle $\chi_1$ as set by the WGC then translates into a bound on the weak scale. From Eq.(\ref{eq:fermionDMmass}), the bound $m_{\chi_1} \lesssim g M_{\rm Pl}$ can be rewritten as
\begin{equation}
	v^2 \lesssim \frac{2}{y^2} (m_L + m_N)(g M_{\rm Pl} - m_N) \ .
\label{eq:weakscalefermionDM}
\end{equation}
For values of the vector-like fermion masses of order the weak scale, and given that the experimentally allowed values of the Yukawa coupling are $y \ll 1$, it is clear from Eq.(\ref{eq:weakscalefermionDM}) that a bound on $v$ of order the weak scale itself requires a rather accurate cancellation between $g M_{\rm Pl}$ and $m_N$. This is indeed the sign of a fine-tuning, albeit of the technically natural parameters $g$ and $m_N / M_{\rm Pl}$.
To be more precise, using the standard fine-tuning measure already introduced in \ref{sec:scalar}, we find a large sensitivity in the value of $v$, as given by Eq.(\ref{eq:weakscalefermionDM}), with respect to the underlying parameters. For example:
\begin{equation}
	\Delta_{m_N} \approx \frac{2 m_N (m_L + m_N)}{y^2 v^2} \sim \frac{1}{y^2} \gtrsim 10^7 \ ,
\end{equation}
which is indeed $\gg 1$, reflecting the level of accuracy needed in the cancellation of the last factor of Eq.(\ref{eq:weakscalefermionDM}). Thus generating the observed DM relic abundance from standard freeze-out of the super-extremal particle in the fermionic model appears to render the weak scale exquisitely sensitive to the values of underlying parameters.

As in the scalar model, additional considerations imposed by the magnetic WGC can potentially be accommodated while preserving the super-extremal fermion as a DM candidate. The $U(1)_X$ can again be embedded into an $SU(2)_X$ gauge group under which $L, L^c, N,N^c$ are all promoted into doublets. This increases the degeneracy of states in the low-energy theory by appropriate factors of 2, but otherwise leaves the parametrics of freeze-out unaltered. It is also conceivable that the fermionic DM candidate could remain viable even if there is more radical new physics appearing at the magnetic WGC scale $\Lambda$, since the freeze-out process occurs at temperatures somewhat below this scale.

\subsection{Plasma effects}
\label{sec:plasma}

The two models discussed in previous sections share a common feature: in both, the DM is charged under a hidden Abelian gauge group with charge $g \lll 1$.
In fact, one might think that for the values of $g$ considered here there is probably no way of probing the $U(1)_X$ charge of the DM, effectively leaving a WGC-based solution of the electroweak hierarchy problem in the dark as far as experimental exploration is concerned.

A caveat to the above expectation concerns the plasma nature of the DM models under consideration.
In all the scenarios considered here, the $2 \rightarrow 2$ scattering rate of DM particles falls below the current Hubble scale, thus making the DM collisionless at short-distances. The DM, however, is made of equal populations of particles and anti-particles carrying opposite $U(1)_X$ charges, and therefore subject to long-range interactions.
In this situation, collective effects mediated by the massless vector dominate over the effect of short-range collisions. On very large scales, such as those encountered in astrophysical settings,  collective plasma instabilities might lead to observable effects, even for exceedingly small values of the gauge coupling. \footnote{We thank Robert Lasenby for valuable discussions on this point.}
This observation was already made in \cite{Ackerman:mha,JeremyMardon}, with \cite{Heikinheimo:2015kra,Heikinheimo:2017meg,Sepp:2016tfs} further noticing that a subcomponent of the DM being a plasma might provide the necessary amount of dissipative dynamics to explain some of the features observed in the DM distribution seen in the Abell 520 cluster collision \cite{Mahdavi:2007yp,Jee:2014hja}.
In the remainder of this section we make an estimate of the length and timescales over which plasma instabilities can grow and result in potentially observable effects, in the context of models where the DM particle approximately saturates the WGC bound. For concreteness, we take the parameters relevant to the Abell 520 cluster as a benchmark.

The characteristic frequency of fluctuations in a plasma is given by the so-called plasma frequency $\omega_p = \sqrt{g^2 n / m}$, with $n$ the number density of the particles making the plasma and $m$ their mass.
It will be convenient to write this in terms of the energy density $\rho = m n$, so that $\omega_p = \sqrt{g^2 \rho / m^2}$.
For a particle with mass $m$ that saturates the WGC bound, $m \lesssim g M_{\rm Pl}$, the corresponding plasma frequency can thus be written as
\begin{equation}
	\omega_p = \sqrt{\frac{g^2 \rho}{m^2}} \gtrsim \frac{\sqrt{\rho}}{M_{\rm Pl}} \ .
	\label{eq:omegapWGC}
\end{equation}

In the context of plasma DM, the value of $\rho$ in Eq.(\ref{eq:omegapWGC}) will vary depending on the astrophysical object under consideration.
For instance, the colliding DM halos inside the Abell 520 cluster have typical sizes $R \approx 200 \ {\rm kpc}$, and masses $M \approx 4 \cdot 10^{13} M_\odot$ \cite{Jee:2014hja}, leading to
$\rho \approx 0.04 \  {\rm GeV} \ {\rm cm}^{-3}$. The typical fluctuation timescale is therefore
\begin{equation}
	\omega_p^{-1} \lesssim 10^{15} \ {\rm s} \ \left( \frac{0.04 \  {\rm GeV} \ {\rm cm}^{-3}}{\rho} \right)^{1/2} \ .
\label{eq:taup}
\end{equation}
For comparison, the typical timescale of a galaxy cluster collision is of order $1 \ {\rm Gyr} \sim 10^{16} \ {\rm s}$.

Small fluctuations might become unstable when two plasma clouds collide. If the two clouds are travelling non-relativistically, as it is the case for two colliding clusters, the fastest growing instability is the electrostatic `two-stream' instability \cite{Bret:2009fy,Dieckmann:2017rty,Dieckmann:2017qwi}. Its maximum growth rate is $\delta \sim \omega_p$, corresponding to a mode with wavelength $\lambda_* \approx 2 \pi v_{col} / \omega_p$, with $v_{col}$ the collision speed \cite{Dieckmann:2017qwi}.
Once the size of the overlapping region becomes of order $\lambda_*$, the instability can start to grow, first exponentially fast, until saturation is achieved. After saturation, an electrostatic shock wave can form, leading to dissipative dynamics and therefore making the DM effectively collisional on very large scales.

The typical speed in a galaxy cluster collision is $v_{col} \sim 10^{-2} $, leading to $\lambda_* \sim 100 \ {\rm kpc}$. This is comparable to the size of the colliding DM halos, thus making it potentially possible for the instability to grow.
Numerical \cite{Dieckmann:2017rty,Dieckmann:2017qwi} and analytical \cite{Dieckmann:2017qwi} studies suggest that it might take a time of order  $\tau_s \sim 10 \omega_p^{-1}$ for the instability to saturate, and a similar timescale for a shock wave to finally develop. Overall, it is possible that after a time $\sim (10 - 10^2)  \omega_p^{-1}$ the effects of the initial fluctuations might have developed into a genuine shock. For the numerical values considered here (see Eq.(\ref{eq:taup})) it is thus possible that such a timescale might be comparable to the timescale of a cluster collision.

A more conclusive statement would require a more refined analysis, which is beyond the scope of this work. However, the possibility that plasma effects might have observables consequences even in models where the DM saturates the WGC bound, regardless of its connection to the weak scale, provides further motivation for dedicated analytical and numerical studies.

\section{Conclusion}

Under generic assumptions about the properties of the UV completion of the SM, it is a mystery that the weak scale appears to be parametrically below much higher scales at which we expect new physics to appear. For a long time, and for compelling reasons, the standard approach to the hierarchy problem has been to try and lower the cutoff of the UV completion close to the weak scale itself, the canonical example being supersymmetric models, with the scale of SUSY-breaking not far above the Higgs vev. This approach has two virtues. First, that it provides an explanation of the smallness of $v$ by means of new symmetries and dynamics, while at the same time allowing us to remain largely agnostic about the specific details of the theory in the far UV. Second, that it necessarily implies new physics not far above the weak scale, making it possible to experimentally falsify specific models' predictions.

However, with increasingly stringent upper bounds on traditional models of naturalness, it seems timely to seriously consider an alternative approach: that the SM may be valid up to scales parametrically above $v$, with a UV completion featuring rather unusual properties far from what, in our ignorance, we might consider generic. The challenge of this approach is to render it concrete. In this paper, we have taken a step towards making this suggestion more tangible by exploring how the WGC might provide an explanation for the smallness of the weak scale. By demanding that the electric WGC for a new Abelian gauge group be satisfied by a particle acquiring at least some of its mass from electroweak symmetry breaking, we have advanced the possibility (first explored in \cite{Cheung:2014vva}) that the weak scale may \emph{naturally} lie parametrically below the UV cutoff. Beyond the presence of an extra Abelian gauge group with gauge coupling smaller than $10^{-16}$, the extra matter content required to link the weak scale to the WGC is fairly minimal. The primary challenge is to address the implications of the magnetic WGC, which imply additional physics entering near the scale of the super-extremal particle. Relative to previous work in this direction \cite{Cheung:2014vva}, our main innovations have been to raise the scale of the magnetic WGC to the weak scale itself; demonstrate the extent to which physics associated with the magnetic WGC can be decoupled from the cutoff of the SM Higgs sector; and highlight the possibility that new physics associated with the magnetic WGC could trivialize the desired bound. In addition, we have explored the possibility of separating the scales of the electric and magnetic WGC via the scalar modification of the conjecture (which would further protect a bound on the weak scale from being trivialized by the magnetic WGC), constructing concrete models that both realize this possibility \cite{Lust:2017wrl} and illustrate its limitations when applied to the weak scale.

The phenomenological implications of a WGC-based solution to the hierarchy problem are not as obvious as those in models of supersymmetry or compositeness. Yet signatures remain at or below the weak scale, making this a productive avenue for experimental tests. We have identified two specific classes of signatures that could provide support for this scenario, namely modifications to the SM Higgs sector and collective plasma effects if the particle satisfying the WGC is itself the DM. Indeed, the distribution of DM in cluster collisions can provide a stringent probe of very tiny gauge couplings if the DM is a plasma, regardless of the role of the DM particle with respect to the hierarchy problem.

There are numerous directions for future development. Clearly, the WGC-based explanations for $v \lll M_{\rm Pl}$ explored here would be more compelling if made in the context of a fully calculable theory where the natural size for $v$ in the absence of the WGC is much above its experimentally measured value -- such as e.g.~in Split Supersymmetry. Embedding the mechanism explored here into such a UV completion would provide a more complete picture of both why electroweak symmetry is broken, and why the weak scale is so small. This work also motivates continued sharpening of both the electric and magnetic forms of the conjecture, as the implications of the magnetic WGC for a potential bound on the weak scale depend sensitively on both the detailed form of each conjecture and the specific way in which the magnetic version is satisfied. The WGC with scalar fields merits comparable attention, given its potential role in protecting a bound against the magnetic WGC. More broadly, although the path to explaining the weak scale by the weakness of gravity appears narrow, it remains an open and inviting avenue for further exploration. 
	
	\section*{Acknowledgments}
	We thank Nima Arkani-Hamed, Victor Gorbenko, Robert Lasenby, John March-Russell, and Eran Palti for helpful conversations. The research of NC and SK is supported in part by the US Department of Energy under the Early Career Award DE-SC0014129 and the Cottrell Scholar Program through the Research Corporation for Science Advancement. The research of IGG is funded by the Gordon and Betty Moore Foundation through Grant GBMF7392.  Research at KITP is supported in part by the National Science Foundation under Grant No.~NSF PHY-1748958.
	
	\bibliography{weakweak}
	
\end{document}